\documentclass[superscriptaddress,twocolumn, prc,showpacs]{revtex4}

\usepackage{graphicx}
\usepackage{epstopdf}
\usepackage{subfigure}
\usepackage{ctable}
\usepackage{siunitx}

\usepackage{float}

\begin{document}

\title{Beta-delayed proton emission in the $^{100}$Sn region}

\author{G.~Lorusso}
\altaffiliation[Current address: ]{RIKEN Nishina Center, 2-1 Hirosawa, Wako, Saitama 351-0198, Japan}
\affiliation{National Superconducting Cyclotron Laboratory, Michigan State University, East Lansing, Michigan 48824, USA}
\affiliation{Joint Institute for Nuclear Astrophysics, Michigan State University, East Lansing, Michigan 48824, USA}
\affiliation{Department of Physics and Astronomy, Michigan State University, East Lansing, Michigan 48824, USA}

\author{A.~Becerril}
\affiliation{National Superconducting Cyclotron Laboratory, Michigan State University, East Lansing, Michigan 48824, USA}
\affiliation{Joint Institute for Nuclear Astrophysics, Michigan State University, East Lansing, Michigan 48824, USA}
\affiliation{Department of Physics and Astronomy, Michigan State University, East Lansing, Michigan 48824, USA}

\author{ A.~Amthor}
\affiliation{National Superconducting Cyclotron Laboratory, Michigan State University, East Lansing, Michigan 48824, USA}
\affiliation{Joint Institute for Nuclear Astrophysics, Michigan State University, East Lansing, Michigan 48824, USA}
\affiliation{Department of Physics and Astronomy, Michigan State University, East Lansing, Michigan 48824, USA}

\author{T.~Baumann}
\affiliation{National Superconducting Cyclotron Laboratory, Michigan State University, East Lansing, Michigan 48824, USA}

\author{D.~Bazin}
\affiliation{National Superconducting Cyclotron Laboratory, Michigan State University, East Lansing, Michigan 48824, USA}

\author{J.S.~Berryman}
\affiliation{National Superconducting Cyclotron Laboratory, Michigan State University, East Lansing, Michigan 48824, USA}
\affiliation{Department of Chemistry, Michigan State University, East Lansing, Michigan 48824, USA}

\author{B.A.~Brown}
\affiliation{National Superconducting Cyclotron Laboratory, Michigan State University, East Lansing, Michigan 48824, USA}
\affiliation{Joint Institute for Nuclear Astrophysics, Michigan State University, East Lansing, Michigan 48824, USA}
\affiliation{Department of Physics and Astronomy, Michigan State University, East Lansing, Michigan 48824, USA}

\author{R.H.~Cyburt}
\affiliation{National Superconducting Cyclotron Laboratory, Michigan State University, East Lansing, Michigan 48824, USA}
\affiliation{Joint Institute for Nuclear Astrophysics, Michigan State University, East Lansing, Michigan 48824, USA}

\author{H.L.~Crawford}
\affiliation{National Superconducting Cyclotron Laboratory, Michigan State University, East Lansing, Michigan 48824, USA}
\affiliation{Department of Chemistry, Michigan State University, East Lansing, Michigan 48824, USA}

\author{A.~Estrade}
\affiliation{National Superconducting Cyclotron Laboratory, Michigan State University, East Lansing, Michigan 48824, USA}
\affiliation{Joint Institute for Nuclear Astrophysics, Michigan State University, East Lansing, Michigan 48824, USA}
\affiliation{Department of Physics and Astronomy, Michigan State University, East Lansing, Michigan 48824, USA}

\author{A.~Gade}
\affiliation{National Superconducting Cyclotron Laboratory, Michigan State University, East Lansing, Michigan 48824, USA}
\affiliation{Department of Physics and Astronomy, Michigan State University, East Lansing, Michigan 48824, USA}

\author{T.~Ginter}
\affiliation{National Superconducting Cyclotron Laboratory, Michigan State University, East Lansing, Michigan 48824, USA}

\author{C.J.~Guess}
\altaffiliation[Current address: ]{Department of Physics and Applied Physics, University of Massachusetts Lowell, Lowell, MA 01854, USA}
\affiliation{National Superconducting Cyclotron Laboratory, Michigan State University, East Lansing, Michigan 48824, USA}
\affiliation{Joint Institute for Nuclear Astrophysics, Michigan State University, East Lansing, Michigan 48824, USA}
\affiliation{Department of Physics and Astronomy, Michigan State University, East Lansing, Michigan 48824, USA}

\author{M.~Hausmann}
\affiliation{Facility for Rare Isotope Beams, Michigan State University, 640 South Shaw Lane, East Lansing, MI 48824, USA}

\author{G.W.~Hitt}
\affiliation{Department of Applied Mathematics and Science, Khalifa University of Science, Technology, and Research, Abu Dhabi Campus, P.O. Box 127788, Abu Dhabi, UAE}

\author{P.F.~Mantica}
\affiliation{National Superconducting Cyclotron Laboratory, Michigan State University, East Lansing, Michigan 48824, USA}
\affiliation{Department of Chemistry, Michigan State University, East Lansing, Michigan 48824, USA}

\author{ M.~Matos}
\affiliation{National Superconducting Cyclotron Laboratory, Michigan State University, East Lansing, Michigan 48824, USA}
\affiliation{Joint Institute for Nuclear Astrophysics, Michigan State University, East Lansing, Michigan 48824, USA}

\author{ R.~Meharchand}
\altaffiliation[Current address: ]{Los Alamos National Laboratory, Los Alamos, NM 87545, USA}
\affiliation{National Superconducting Cyclotron Laboratory, Michigan State University, East Lansing, Michigan 48824, USA}
\affiliation{Joint Institute for Nuclear Astrophysics, Michigan State University, East Lansing, Michigan 48824, USA}
\affiliation{Department of Physics and Astronomy, Michigan State University, East Lansing, Michigan 48824, USA}

\author{K.~Minamisono}
\affiliation{National Superconducting Cyclotron Laboratory, Michigan State University, East Lansing, Michigan 48824, USA}

\author{F.~Montes}
\affiliation{National Superconducting Cyclotron Laboratory, Michigan State University, East Lansing, Michigan 48824, USA}
\affiliation{Joint Institute for Nuclear Astrophysics, Michigan State University, East Lansing, Michigan 48824, USA}

\author{ G.~Perdikakis}
\affiliation{National Superconducting Cyclotron Laboratory, Michigan State University, East Lansing, Michigan 48824, USA}
\affiliation{Joint Institute for Nuclear Astrophysics, Michigan State University, East Lansing, Michigan 48824, USA}

\author{J.~Pereira}
\affiliation{National Superconducting Cyclotron Laboratory, Michigan State University, East Lansing, Michigan 48824, USA}
\affiliation{Joint Institute for Nuclear Astrophysics, Michigan State University, East Lansing, Michigan 48824, USA}

\author{M.~Portillo}
\affiliation{National Superconducting Cyclotron Laboratory, Michigan State University, East Lansing, Michigan 48824, USA}

\author{H. Schatz}
\affiliation{National Superconducting Cyclotron Laboratory, Michigan State University, East Lansing, Michigan 48824, USA}
\affiliation{Joint Institute for Nuclear Astrophysics, Michigan State University, East Lansing, Michigan 48824, USA}
\affiliation{Department of Physics and Astronomy, Michigan State University, East Lansing, Michigan 48824, USA}

\author{K. Smith}
\affiliation{National Superconducting Cyclotron Laboratory, Michigan State University, East Lansing, Michigan 48824, USA}
\affiliation{Joint Institute for Nuclear Astrophysics, Michigan State University, East Lansing, Michigan 48824, USA}
\affiliation{Department of Physics and Astronomy, Michigan State University, East Lansing, Michigan 48824, USA}

\author{ J.~Stoker}
\affiliation{National Superconducting Cyclotron Laboratory, Michigan State University, East Lansing, Michigan 48824, USA}
\affiliation{Department of Chemistry, Michigan State University, East Lansing, Michigan 48824, USA}

\author{A.~Stolz}
\affiliation{National Superconducting Cyclotron Laboratory, Michigan State University, East Lansing, Michigan 48824, USA}

\author{R.G.T.~Zegers}
\affiliation{National Superconducting Cyclotron Laboratory, Michigan State University, East Lansing, Michigan 48824, USA}
\affiliation{Joint Institute for Nuclear Astrophysics, Michigan State University, East Lansing, Michigan 48824, USA}
\affiliation{Department of Physics and Astronomy, Michigan State University, East Lansing, Michigan 48824, USA}

\date{\today}

\begin{abstract}
$\beta$-delayed proton emission from nuclides in the neighborhood of $^{100}$Sn was studied at the National Superconducting Cyclotron Laboratory. The nuclei were produced by fragmentation of a 120~MeV/nucleon $^{112}$Sn primary beam on a Be target. Beam purification was provided by the A1900 Fragment Separator and the Radio Frequency Fragment Separator. The fragments of interest were identified and their decay was studied with the NSCL Beta Counting System (BCS) in conjunction with the Segmented Germanium Array (SeGA).
The nuclei $^{96}$Cd, $^{98}$In$^{g}$, $^{98}$In$^{m}$ and $^{99}$In were identified as $\beta$-delayed proton emitters, with branching ratios $b_{\beta p}=5.5(40)\%$, 5.5$^{+3}_{-2}$\%, 19(2)\% and 0.9(4)\%, respectively. The $b_{\beta p}$ for $^{89}$Ru, $^{91,92}$Rh, $^{93}$Pd and $^{95}$Ag were deduced for the first time with $b_{\beta p}= 3^{+1.9}_{-1.7}\%$, 1.3(5)\%, 1.9(1)\%, 7.5(5)\% and 2.5(3)\%, respectively. The $b_{\beta p}=22(1)\%$ for $^{101}$Sn was deduced with higher precision than previously reported. The impact of the newly measured $b_{\beta p}$ values on the composition of the type-I X-ray burst ashes was studied.

\end{abstract}


\maketitle

\section{Introduction}

Nuclei located around the doubly-magic nucleus $^{100}$Sn offer a unique testing ground to improve the understanding of nuclear structure far from stability, as the proximity to a doubly-magic core makes shell-model calculations feasible (see for example Refs.~\cite{Oga83,Schm97,Bat03,Graw05,Muk04,Bec11}). The influence of the high-spin $g_{9/2}$ single-particle orbital on the valence nucleon or hole structure and the proton-neutron interactions enhanced in $N \approx Z$ nuclei \cite{Eng97}, lead to a broad range of spin and seniority effects including isomers \cite{EsZ06}, which can be sensitive probes of nuclear structure. In addition, the large $Q$ values for $\beta$ decay enable the use of a rich set of experimental tools to study these nuclei despite their low production rates. These tools include $\beta$-delayed $\gamma$ and proton spectroscopy and direct proton emission. $\alpha$ decay from the $Z>50$ region offers an additional opportunity to probe nuclear quantum structure with high efficiency and selectivity \cite{Lid06}. In this paper, we report the observation of $\beta$-delayed proton ($\beta p$-) emission for a wide range of nuclei at or close to $N=Z$ and with element numbers just below Sn. We measured $\beta p$-energy spectra as well as $\gamma$-energy spectra in coincidence with $\beta p$ events, and deduced $\beta p$-branching ratios $b_{\beta p}$. New data for the decay of $^{101}$Sn are also presented. The ground-state spin of $^{101}$Sn is of interest to determine the single-particle level ordering as probed by the single neutron outside the $^{100}$Sn core \cite{Sew07,Dar10}.

$\beta p$ emission from nuclei in the vicinity of $^{100}$Sn are also relevant to the understanding of the rapid-proton-capture process (rp process \cite{WaW81}) in type I X-ray bursts \cite{StB06,Scha06}. Such bursts are the most frequent thermonuclear explosions observed in the universe, and are powered by the explosive nuclear burning of hydrogen and helium in a thin layer of fuel accumulated on the surface of an accreting neutron star. Models indicate that, under some conditions, proton captures and $\beta$ decays drive the energy-generating reaction sequence into the $^{100}$Sn region, with the $\alpha$-unbound Te isotopes providing a natural endpoint \cite{Scha01}. Such an extended rp process provides an explanation for bursts that last several minutes.
The lifetimes of the slower $\beta$-decaying isotopes in the rp process, known as waiting points, determine the composition of the burst ashes, which is an important input parameter for neutron star crust models related to a number of observable phenomena \cite{Gup07}. With the recently reported ground state half-lives of $^{96,97}$Cd \cite{Baz08,Lor11} and the improved half-life of $^{84}$Mo~\cite{Stok09}, the ground state half-lives of all waiting points are now known experimentally.

$\beta$-delayed proton emission of nuclei along the rp process path further alters the composition of the burst ashes depending on their $b_{\beta p}$. However, the $b_{\beta p}$ values along the rp-process path are still largely unknown. In this context, the $\beta p$ emission of $^{93}$Pd and $^{97}$Cd are particularly important because, via $\beta p$ emission, they are progenitors of $^{92}$Mo and $^{96}$Ru, whose unusually high abundance in the solar system is not well understood. The study of $^{93}$Pd and $^{97}$Cd is therefore relevant for the understanding of an rp-process contribution to the origin of these isotopes, provided an rp-process site that ejects sufficient amounts of the isotopes is found \cite{Wei06}. The results from our study of the decay of $^{93}$Pd are presented here for the first time, together with a brief discussion on $^{97}$Cd, whose decay properties were already published in Ref.~\cite{Lor11}.

\section{Experimental technique}

The production of $^{100}$Sn and neighboring neutron-deficient isotopes at the NSCL Coupled Cyclotron Facility was described in Ref.~\cite{Baz08}. 
Neutron-deficient nuclides were produced by fragmentation of a 120~MeV/nucleon $^{112}$Sn beam, with an average intensity of 10.7 p-nA, 
impinging upon a 195~mg/cm$^{2}$ ${^9}$Be target. Reaction products were selected with the NSCL A1900 fragment separator \cite{Mor03} operated in achromatic mode. 
A 40.6~mg/cm$^{2}$ thick Kapton wedge, shaped to preserve the achromatism, was placed at the intermediate image of the A1900 to select nuclear charge $Z$. 
The rigidities of the four A1900 dipoles were set to the following values: $B\rho_{1,2}$ = 2.8802~Tm and $B\rho_{3,4}$ = 2.7710~Tm.
The magnetic rigidity selection by the A1900 fragment separator is not sufficient for a decay spectroscopy experiment with neutron-deficient isotopes. The momentum distribution of nuclei produced by projectile fragmentation at intermediate energy (50--200~MeV/nucleon) is asymmetric, and the exotic neutron-deficient species of interest have magnetic rigidities that overlap with the low-momentum tails of more intense fragments closer to stability. 
These more intense fragments have lower nuclear charge but higher velocity than the isotopes of interest resulting in similar energy losses in matter. The resulting fragments have similar $B \rho$ values and cannot be separated by the A1900, therefore additional separation by velocity is needed. The NSCL Radio Frequency Fragment Separator (RFFS) \cite{Baz09} was employed to provide a sinusoidal electric field to deflect ions depending on their phase relationship with the cyclotron frequency. This phase relationship at the downstream location of the RFFS depends on the particle velocity. Particles of interest were selected by a set of vertical slits downstream of the device. 

The magnetic- and velocity-purified beam was then sent to a stack of Si detectors that were part of the NSCL Beta Counting System (BCS)~\cite{Pri02}. After traversing three Si PIN detectors with thicknesses of 297, 297 and 488 \si{\micro \meter}, the beam was implanted into a 985~\si{\micro \meter} tick double-sided silicon strip detector (DSSD). The DSSD was segmented into 40 1-mm wide strips horizontally and vertically, resulting in 1600 pixels. The beam was defocussed over the surface of the DSSD to minimize the implantation rate per pixel. The DSSD was read out with dual gain pre-amplifiers, with the low-gain signal (up to 3 GeV) providing the location and time of ion implantation and the high-gain signal (up to 100~MeV) used to measure position, time and energy deposited by emitted protons and positrons. The DSSD was followed by a $\beta$ calorimeter consisting of six single-sided silicon strip detectors (SSSD) approximately 1~mm thick, and a planar Ge detector approximately 10~mm thick. The calorimeter was capable of stopping $\beta$ particles with energies up to 14~MeV. The calorimeter detectors also served as a veto for light particles transmitted along with the secondary beam, and whose energy loss in the DSSD was comparable to the energy deposited by a $\beta$ particle. These light ions would otherwise contribute to $\beta$ background. $\gamma$ rays emitted within 20~\si{\micro}s of either an implanted ion, a proton, or a positron were detected with 16 high purity germanium detectors from the NSCL Segmented Germanium Array (SeGA)~\cite{Mue01}. The detectors were arranged in two concentric rings, one upstream and one downstream of the DSSD, resulting in an efficiency of about 6\% at 1~MeV and an energy resolution ranging from 2.5 to 2.8~keV FWHM. 

\begin{figure}[h]
\centering
\includegraphics[angle=0,width=0.45\textwidth]{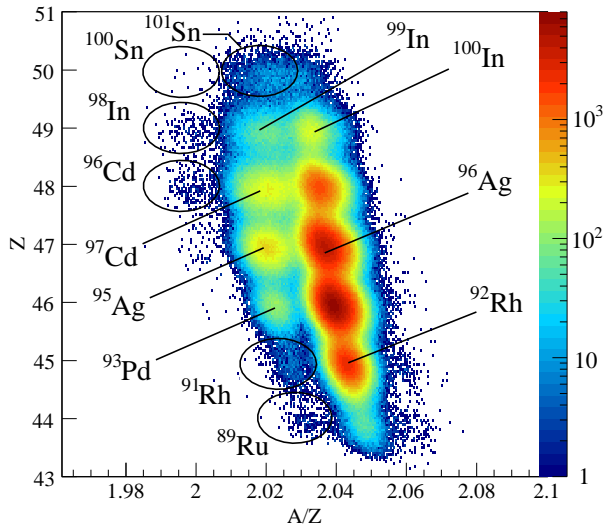}
\vspace{0mm}
\caption{(color online). Particle identification spectrum of the nuclei transmitted through the RFFS. The group of low-Z contaminants that passed through the RFFS at a phase difference of around 360$^0$ is not shown. The species labeled are discussed in the text or have measured properties reported in Table.~\ref{tab_results}.}
\label{PID}
\end{figure}

The three PIN detectors upstream of the DSSD provided redundant event-by-event identification of the incoming particles via energy loss and time-of-flight relative to a start signal provided by a plastic scintillator in the A1900 focal plane. The momentum acceptance of the A1900 was limited to 1\% to allow a separation of isotopes via time-of-flight. The intensity of the secondary beam at the dispersive focal plane of the A1900 was too high (larger than 2 MHz) to allow a magnetic rigidity measurement of beam particles by position tracking with a plastic scintillator. Such a measurement would have allowed use a larger momentum acceptance. In Fig.~\ref{PID} is shown the particle identification spectrum for the most exotic isotopes. The particle identification was confirmed via the well-known $\gamma$ radiation from the deexcitation of \si{\micro}s isomers in $^{90}$Mo, $^{93}$Ru, $^{94,96}$Pd, $^{98}$Cd. These \si{\micro}s isomers, besides $^{94}$Pd and $^{98}$Cd, were transmitted only when the RFFS slits were opened in a dedicated run at the beginning of the experiment. Production cross sections for the $N=Z$ nuclei $^{100}$Sn, $^{98}$In, $^{96}$Cd and the $N=Z+1$ nuclei $^{99}$In and $^{97}$Cd were reported in Ref.~\cite{Baz08}.

\subsection{$\beta$ decay} 
\label{bdecay}
Low-energy signals resulting from $\beta$ or $\beta p$ decay produce signatures in the DSSD that are dinstinct from those produced by heavy ion implantation. These events were correlated with preceding ion events using position and time information. The position correlation area was cross shaped, consisting of the detector pixel where the decay event occurred, plus the four nearest neighbors (two horizontal and two vertical).
The detection efficiency for $\beta$ decays correlated correctly to the preceding implantation of the parent ion for this cross-shaped position correlation area was determined to be 36(4)\%. The efficiency was limited by the energy thresholds of $\sim$200~keV, set in software on the high gain pre-amplifier readout to cut electronic noise. The correlation time window for the decay-curve analysis was chosen to be about 20 times larger than the expected half-life. This choice in temporal correlation allowed the best estimate of the background and contributions from longer-lived decay components, such as daughter and granddaughter decays. Characterization of the long-time components in the decay curve significantly improves the precision of the deduced parent half-lives. The correlation time was shortened to less than five half-lives to optimize the signal-to-background ratio when analyzing $\beta$-delayed $\gamma$ radiation. In cases where the correlation of a decay event with a preceding ion implantation was not unique, all possible correlations were considered to avoid time-dependent background.

The total implantation rate over the entire DSSD implantation detector was about 50 counts per second, and the resulting maximun background rate was 0.03 counts per second per implantation pixel. 

The time differences between implantation and decay of a particular nuclide were histogrammed to produce decay curves, which were fitted based on the maximization of a Poisson probability log-likelihood function that considered the exponential decay of the parent as well as contributions of daughter and granddaughter decays. The effect of $\beta p$ emission on $\beta$-decay-daughter half-lives was neglected due to the small branching ratios. A constant background was also included in the fit. The minimum in the Poisson probability function was varied to ensure the final results were robust. 

A prerequisite for the correlation analysis was the proper identification of the implanted ion species. As shown in Fig.~\ref{PID}, the tails of the time-of-flight distributions of neighboring isotopes overlapped. This overlap was accounted for in the analysis of $^{96}$Cd, $^{98}$In and $^{100}$Sn decays by assigning particle-identification probabilities based on the shapes of the time-of-flight distributions \cite{Baz08}. For example, if a minimum probability of 50\% was desired to consider a nucleus as ``uniquely identified", the set of nuclei $^{96}$Cd, $^{98}$In, and $^{100}$Sn would have contamination levels of 13\%, 15\%, and 14\%, respectively. In such cases, an additional exponential decay component, with the half-life of the neighboring $A+1$ isotope and weighted with the calculated contamination level, was added to the decay curve fit function. 

\subsection{$\beta$-delayed proton emission}
Some $\beta$-decay events were accompanied by proton emission, resulting in a higher energy deposition in the DSSD. The determination of branchings for $\beta$-delayed proton emission required the disentanglement of pure $\beta$ events and $\beta p$ events. In Fig.~2(a) is shows the DSSD energy spectrum for the decay of $^{96}$Ag. Energy loss of $\beta$ particles alone, along with electrons from internal conversion, contributed to the lower energy part of the spectrum, while $\beta p$ events comprised the higher-energy part of this spectrum. Clearly, the two distributions overlap, and separation of the $\beta p$ events was accomplished based on energy deposition patterns. $\beta$ particles are not expected to deposit an energy loss of more than 1~MeV in a single detector pixel, while protons in this mass region typically have energies above 1~MeV, owing to the Coulomb barrier, and are expected to deposit nearly all this energy in one pixel [see Fig.~2(b) and Fig.~2(c)]. Therefore, $\beta p$ events were identified by requiring an energy deposition of more than 1 MeV in the implantation pixel or in nearest neighbor pixels [see Fig.~2(d)]. In principle, $\beta p$ decays may be missed if the $\beta p$ event was "shared" by adiacent strips. The deposition of less than 1~MeV in each of two neighboring strips is then only possible for proton energies below 2~MeV. The probability for such events is low because of the small range of low energy protons in Si. MCNPX simulations confirmed that even for protons with energies of 1.5~MeV only 1\% of the events would be missed. 
MCNPX gave similar results regardless of wheather a uniform implantation depth distribution or a Gaussian distribution (FWHM 100 \si{\micro}m, as predicted by the ion transport code LISE++) were used.

The decomposition of the DSSD decay energy spectra into $\beta$ and $\beta p$ events was tested using coincident $\gamma$ spectra for the $\beta p$ precursors $^{97}$Cd, $^{96}$Ag, and $^{93}$Pd. The $\gamma$ spectra gated on $\beta p$ events only contained transitions from the depopulation of states in the $\beta p$ daughter, as expected, with no evidence for $\gamma$ rays from deexcitation of states in the $\beta$-decaying daughter. Likewise, $\gamma$-ray spectra gated on $\beta$ events only contained $\gamma$ transitions from the depopulation of states in the $\beta$ daughter, with no evidence of $\gamma$ rays from the $\beta p$ daughter.
\begin{figure}[ht]
\begin{flushleft}
\centering
\hspace{-7.5mm}
 \subfigure{\includegraphics[width=0.5\textwidth]{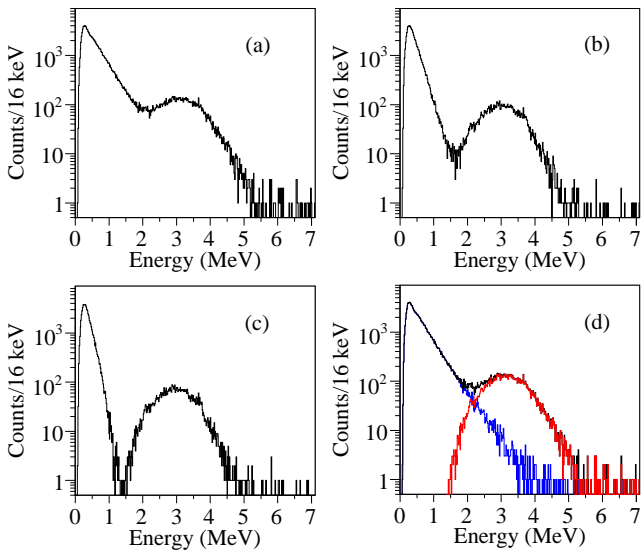}}
 \end{flushleft}
  \caption{(color online). Spectrum of the energy deposited in the DSSD by the decay events of $^{96}$Ag (a) for all events (b) selecting events that deposited energy only in one strip, (c) selecting events that deposited energy only in one pixel, (d) using the selection procedure described in the text, where each $\beta$-decay event (black histogram) is classified as followed by proton emission (red histogram) or not followed by proton emission (blue histogram).}
  \label{pseparation}
\end{figure}

\begin{figure}[h]
\centering
\hspace{-8mm}
 \subfigure{\label{}\includegraphics[width=.22\textwidth]{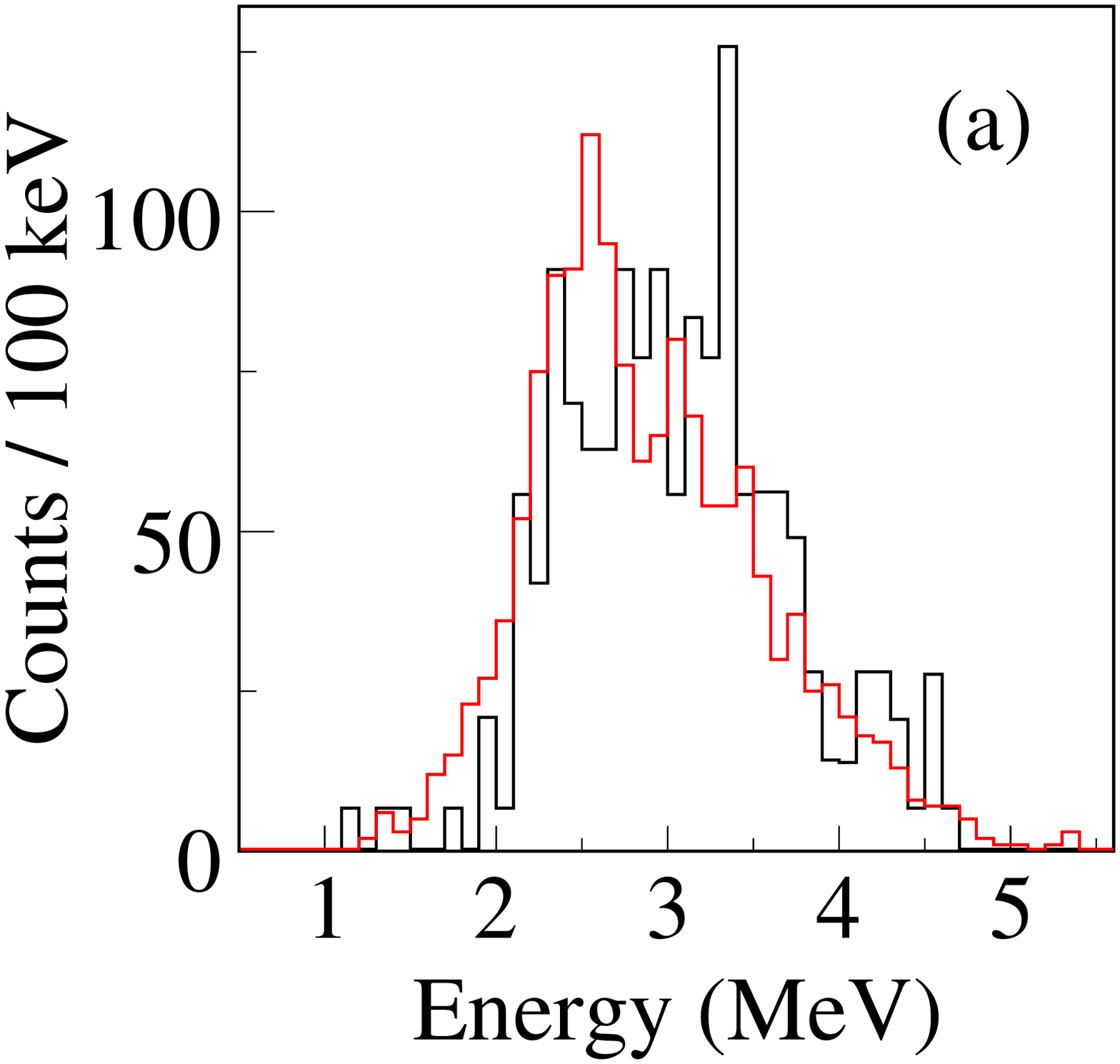}}
 \subfigure{\label{}\includegraphics[width=.22\textwidth]{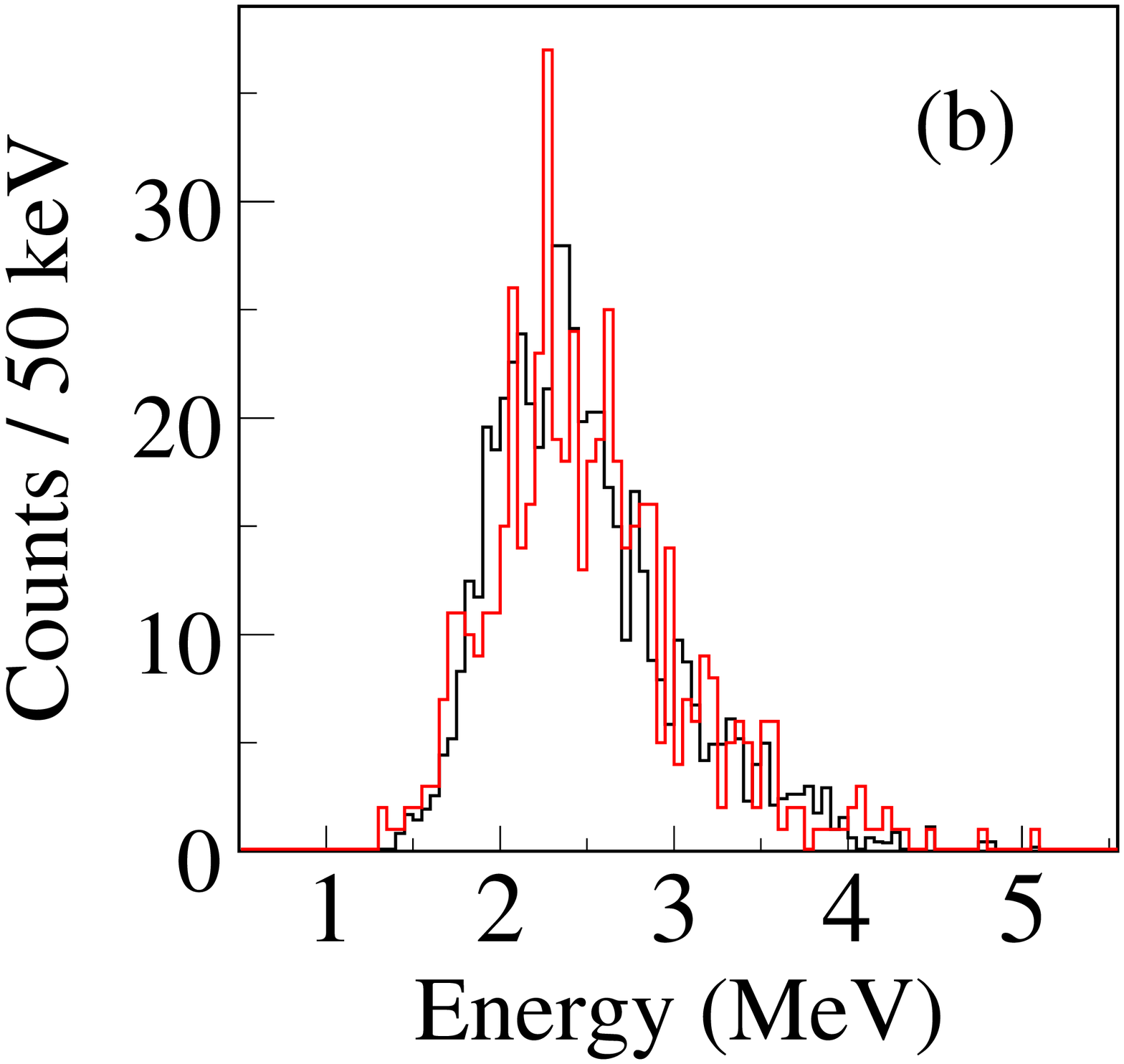}}
 \subfigure{\label{ag96cal1}\includegraphics[width=.22\textwidth]{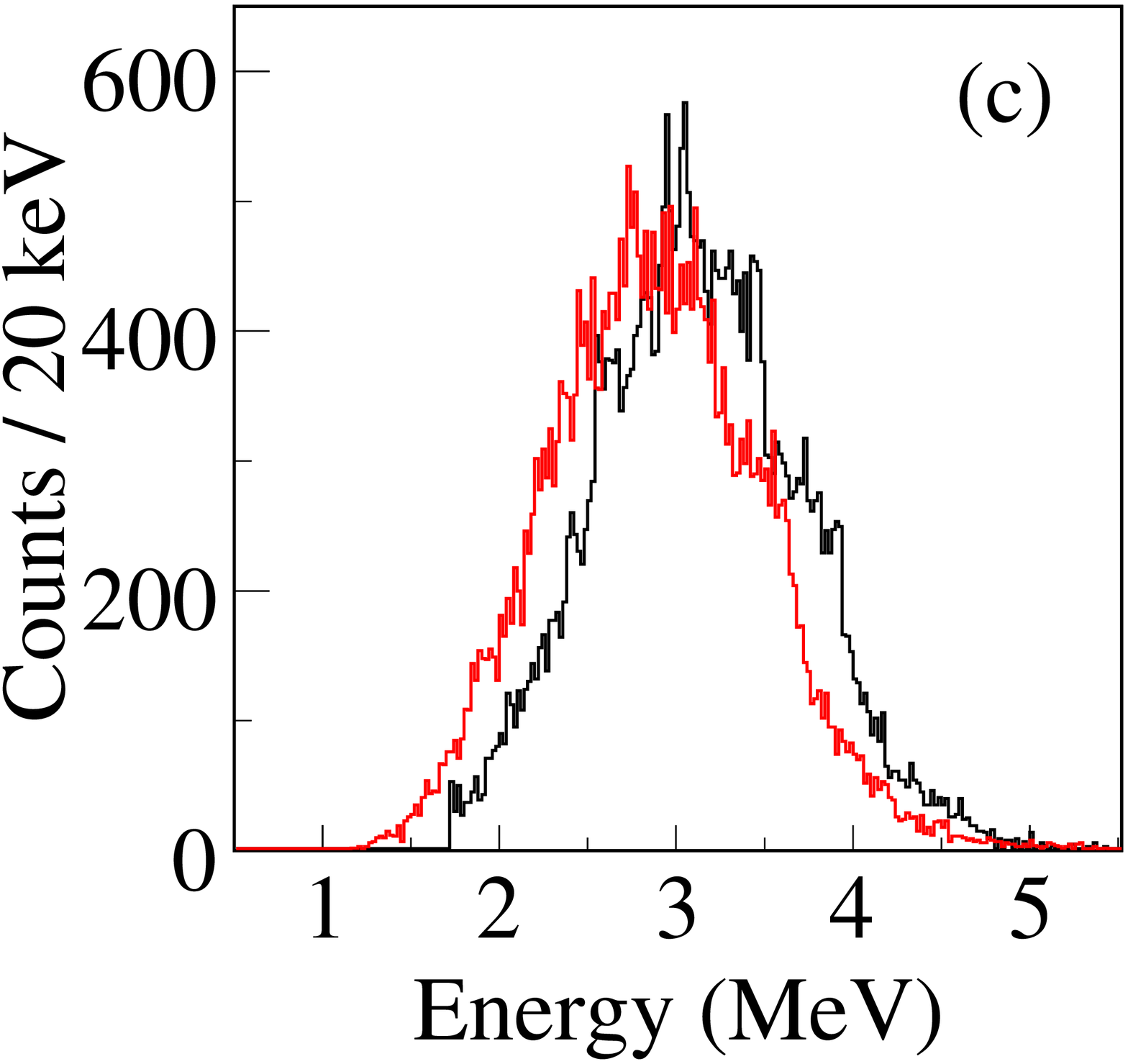}}
\hspace{-2.mm}
 \subfigure{\label{}\includegraphics[width=.22\textwidth]{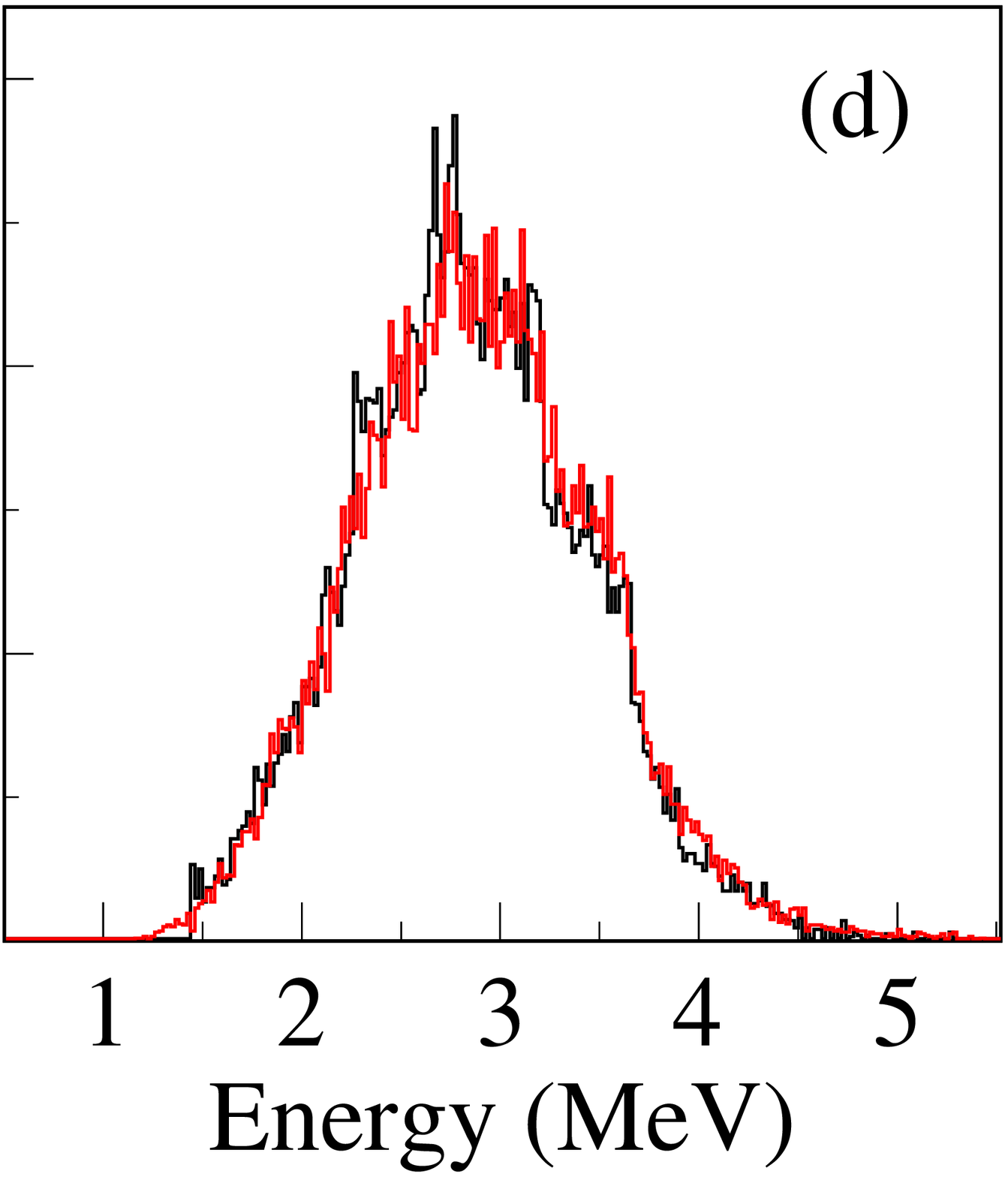}}
 \caption{(color online). Comparison between the calibrated $\beta p$-energy spectra measured in this work (red histogram) and the literature (black histogram) for (a) $^{93}$Pd \cite{Schm00}, (b) $^{95}$Ag \cite{Schm94}, and (c) $^{96}$Ag \cite{Bat03}. In panel (d), data for $^{96}$Ag are displayed after applying a -280~keV energy offset to the literature data.}
 \label{calibration}
\end{figure}

The $\beta p$-energy spectra include a contribution from summing with the simultaneously-emitted $\beta$ particle. The energy in the detection pixel was assigned as the proton energy, assuming that energy deposition in neighboring pixels was mostly due to $\beta$ particles. This assumption was supported by simulations and by the comparison of known proton energy spectra from the literature, as shown in Fig.~\ref{calibration}. The literature spectra were not affected by summing effects, because the relevant experiments were conducted with lower-energy beams, and proton energies were recorded by detectors outside the implantation material.
 The good agreement between the literature and the $\beta p$ spectra from this work suggest that the electron capture ($EC$) branch is significant. Higher-energy excited states in the daughter with energy above $S_p$ are likely fed by $EC$ (see, for example, the $EC$/$\beta^+$ ratio calculated in Refs.~\cite{Zwe49,Wap59}) that does not lead to summing.

An intrinsic $\beta p$-detection efficiency of 100\% was assumed in the analysis presented here, since the proton energy ($>$1~MeV) was well above the high-energy detection threshold ($\sim$200~keV). 
Absolute $\beta p$ branchings were deduced from the number of detected $\beta p$ events correlated with heavy ions implantations, which were corrected for dead-time on a run-by-run basis. The dead-time corrections were typically less than 10\%. The Si detectors of the BCS were energy calibrated using a $^{228}$Th $\alpha$ source. The calibration was linearly extrapolated from the 5~MeV range to lower energies, which introduced a systematic uncertainty of about 100~keV at 1~MeV, the lowest energy of relevance here. The calibrated energy spectra of protons from the decay of $^{96,95}$Ag, and $^{93}$Pd are compared with the spectra reported in the literature in Fig.~\ref{calibration}. Aside from an unexplained offset of 280~keV in the case of $^{96}$Ag [Fig.~\ref{ag96cal1}], the calibrated spectra agree with the literature.

\section{Experimental results}
\label{results}

\begin{figure}[h!]
\vspace{2.6mm}
\begin{center}
  \subfigure{\label{rh92}\includegraphics[width=0.2187\textwidth]{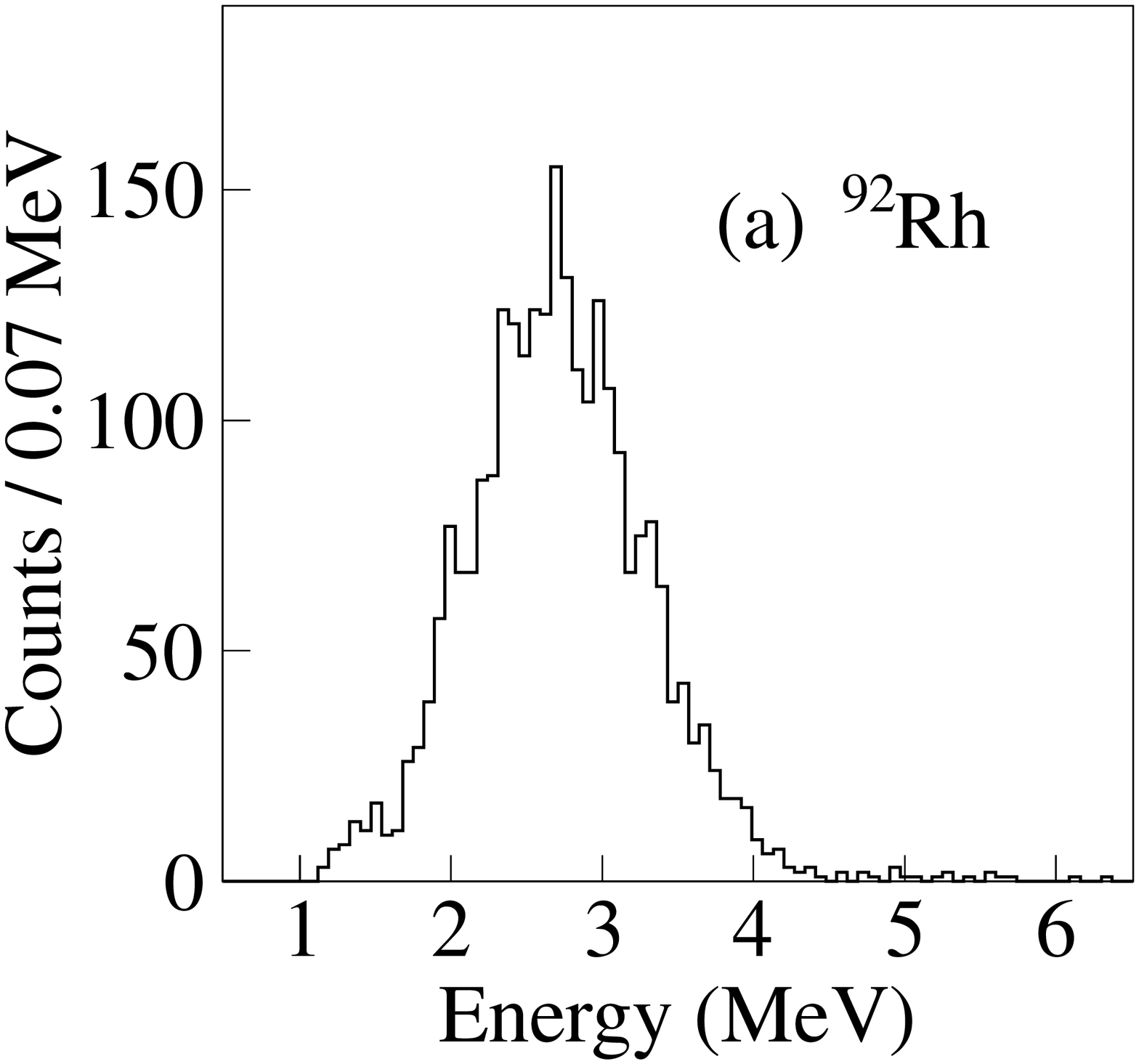}}
\hspace{2mm}
  \subfigure{\label{pd93}\includegraphics[width=0.2187\textwidth]{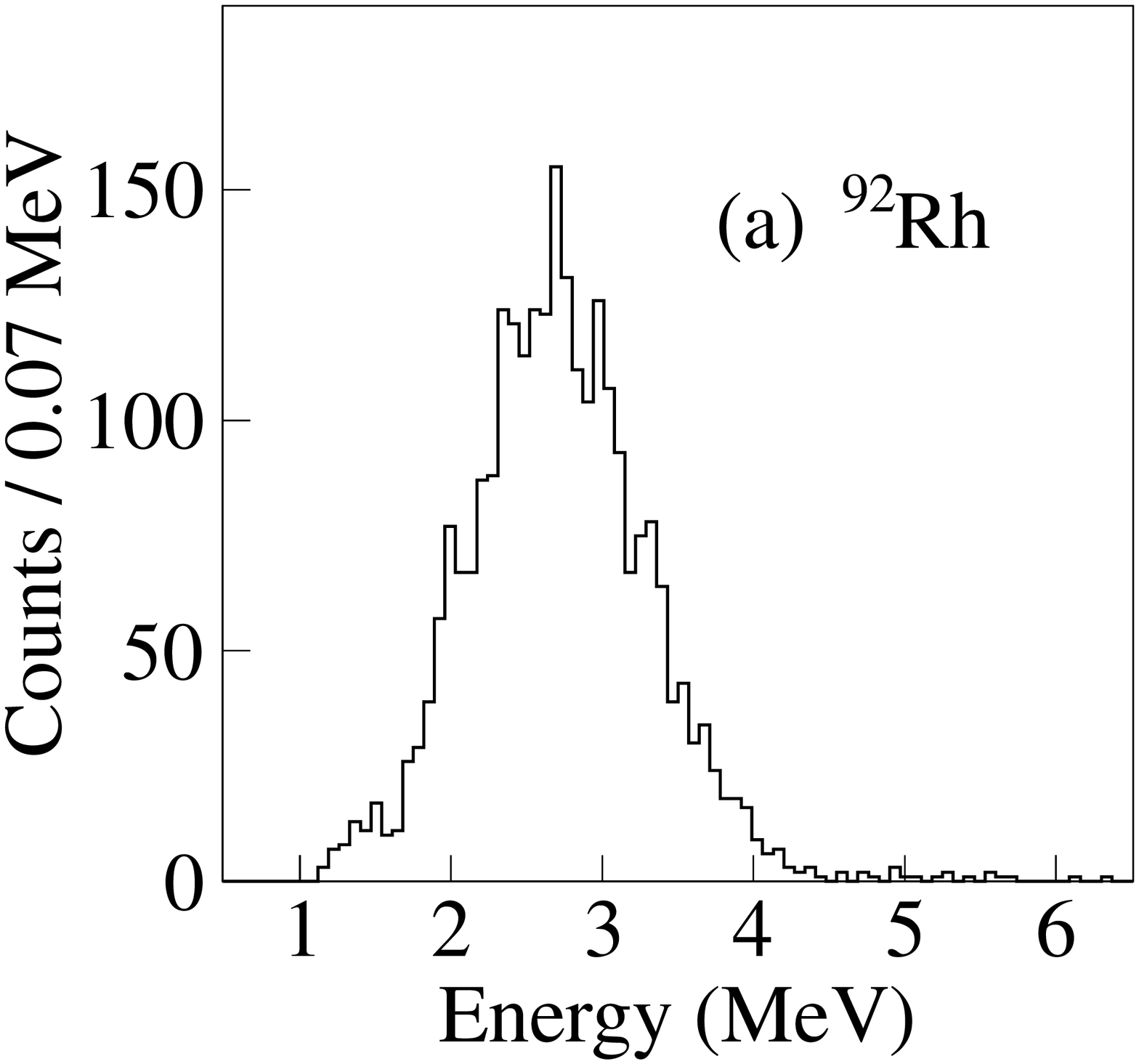}}
\hspace{2mm}
  \subfigure{\label{ag96}\includegraphics[width=0.2187\textwidth]{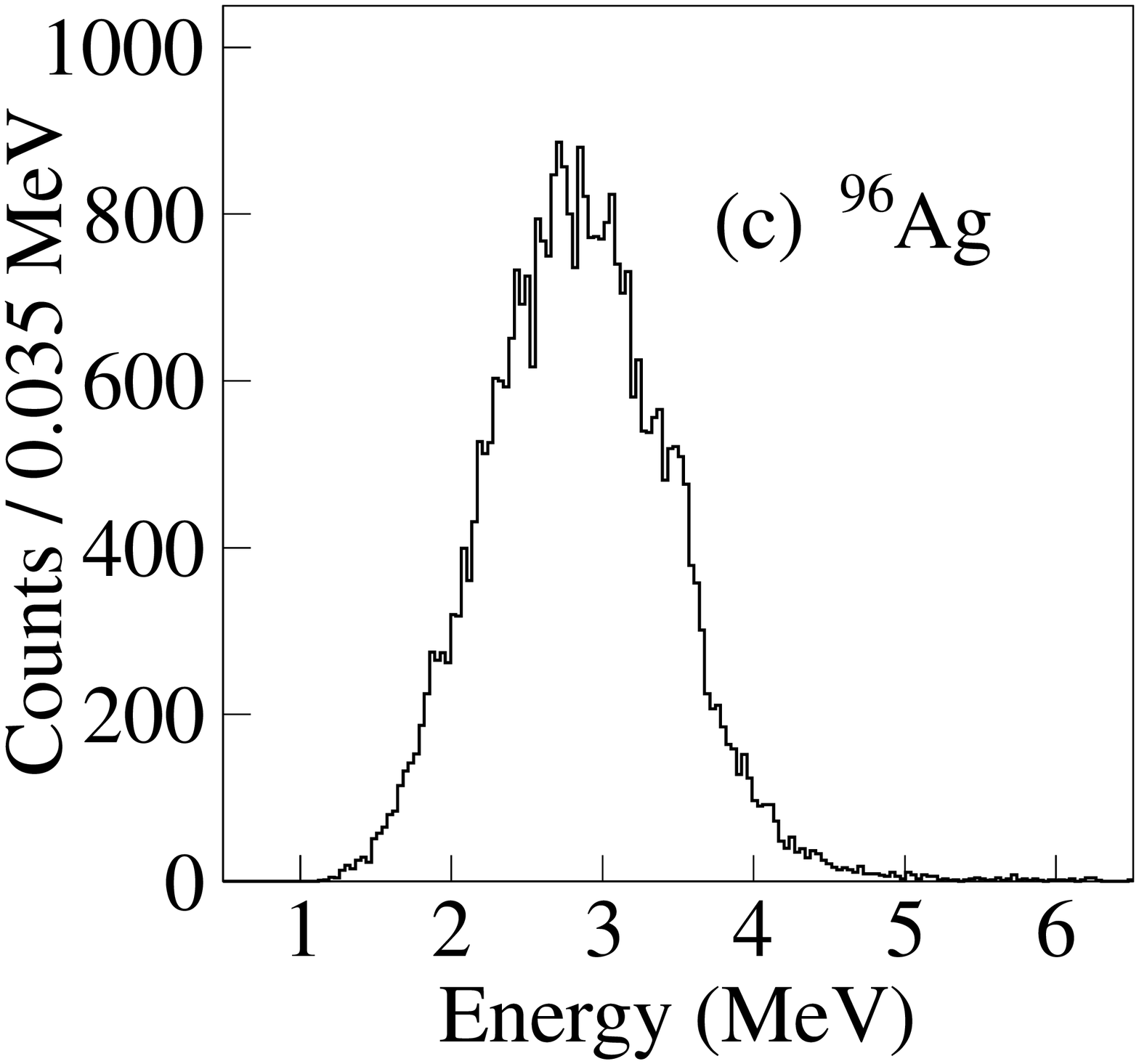}}
\hspace{2mm} 
 \subfigure{\label{ag95}\includegraphics[width=0.2187\textwidth]{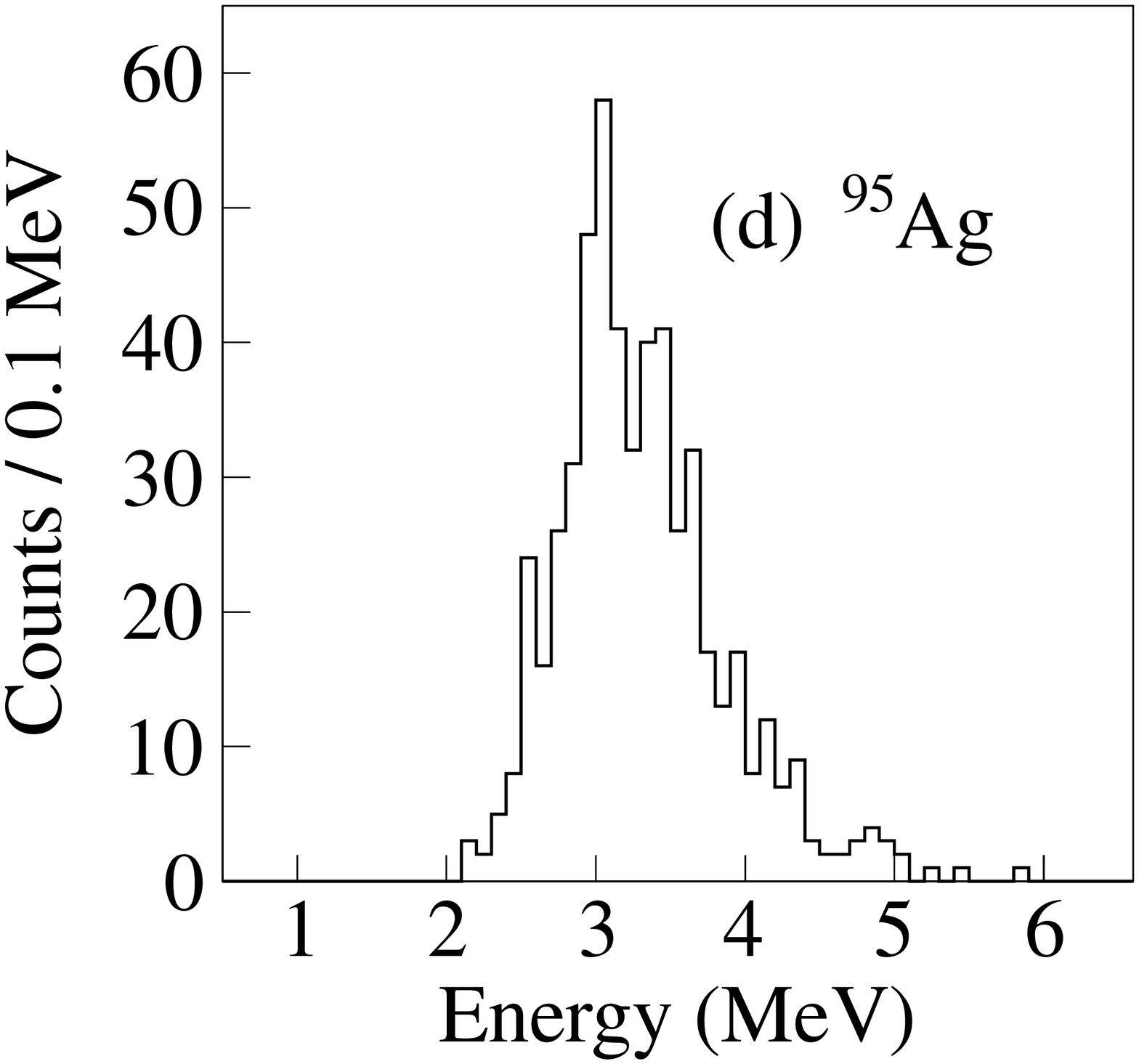}}
\hspace{2mm}
  \subfigure{\label{cd97}\includegraphics[width=0.2187\textwidth]{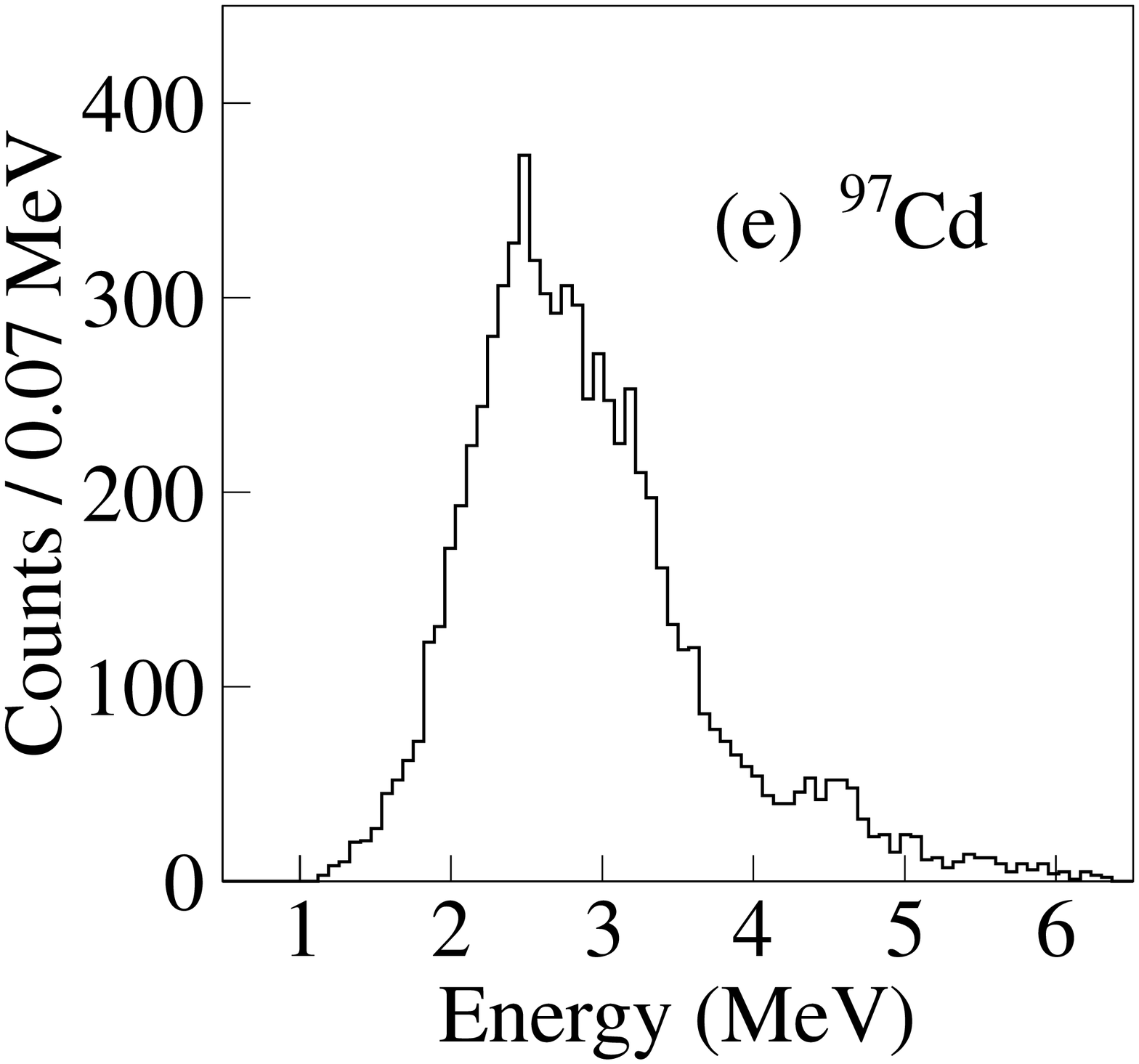}}
\hspace{2mm}
  \subfigure{\label{cd96}\includegraphics[width=0.2187\textwidth]{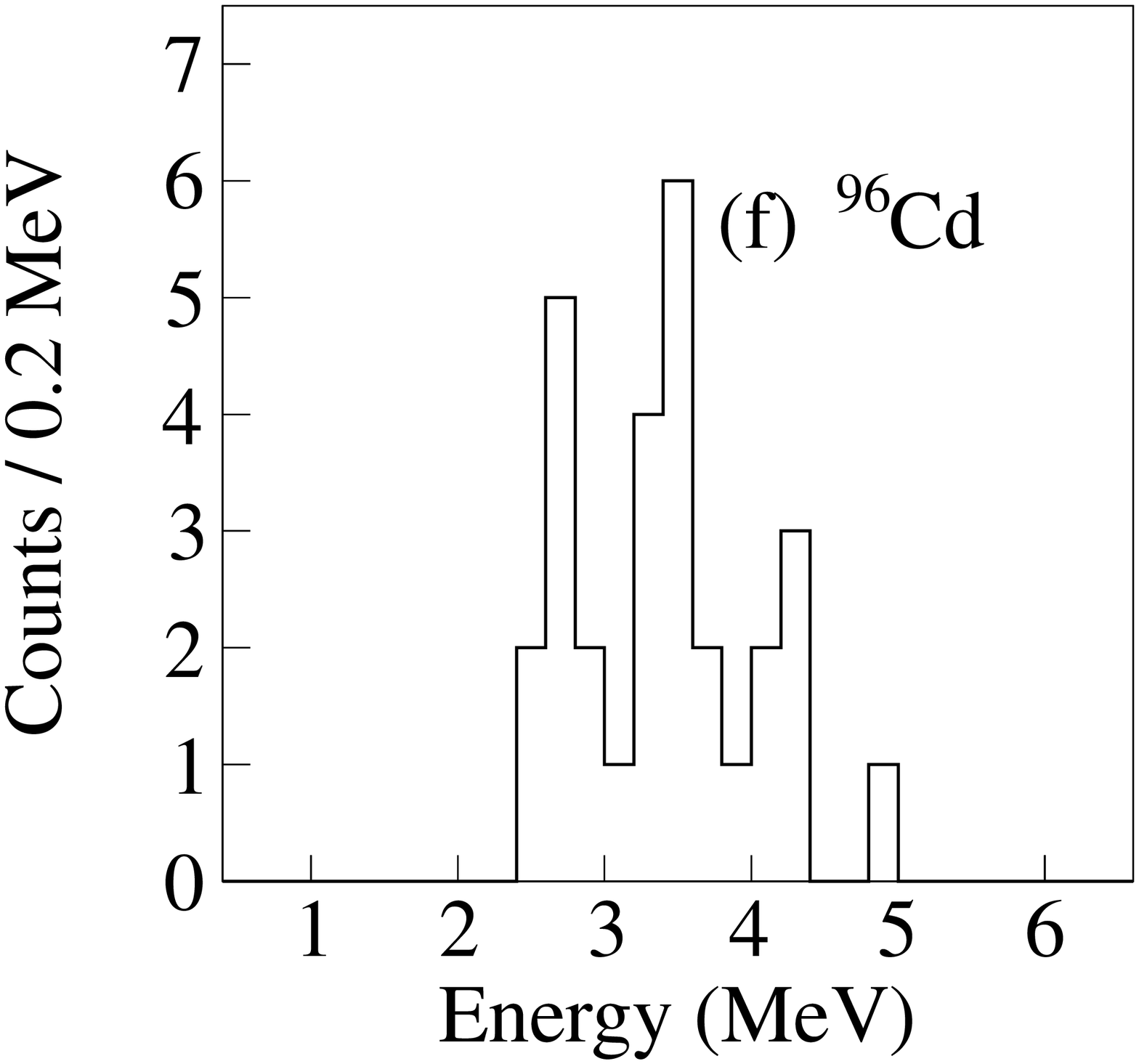}}
\hspace{2mm}
 \subfigure{\label{in100}\includegraphics[width=0.2187\textwidth]{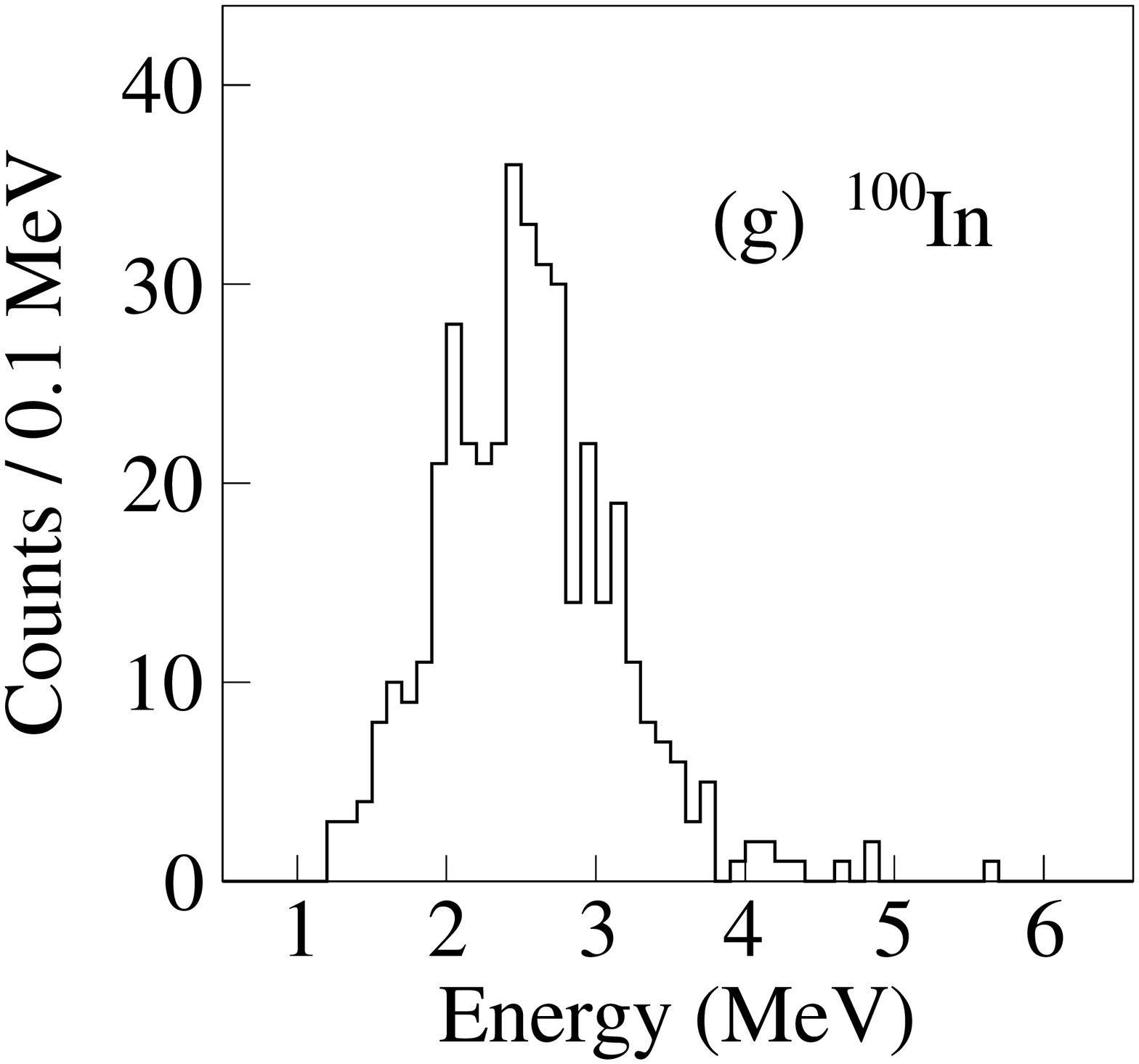}}
\hspace{2mm}
 \subfigure{\label{in99}\includegraphics[width=0.2187\textwidth]{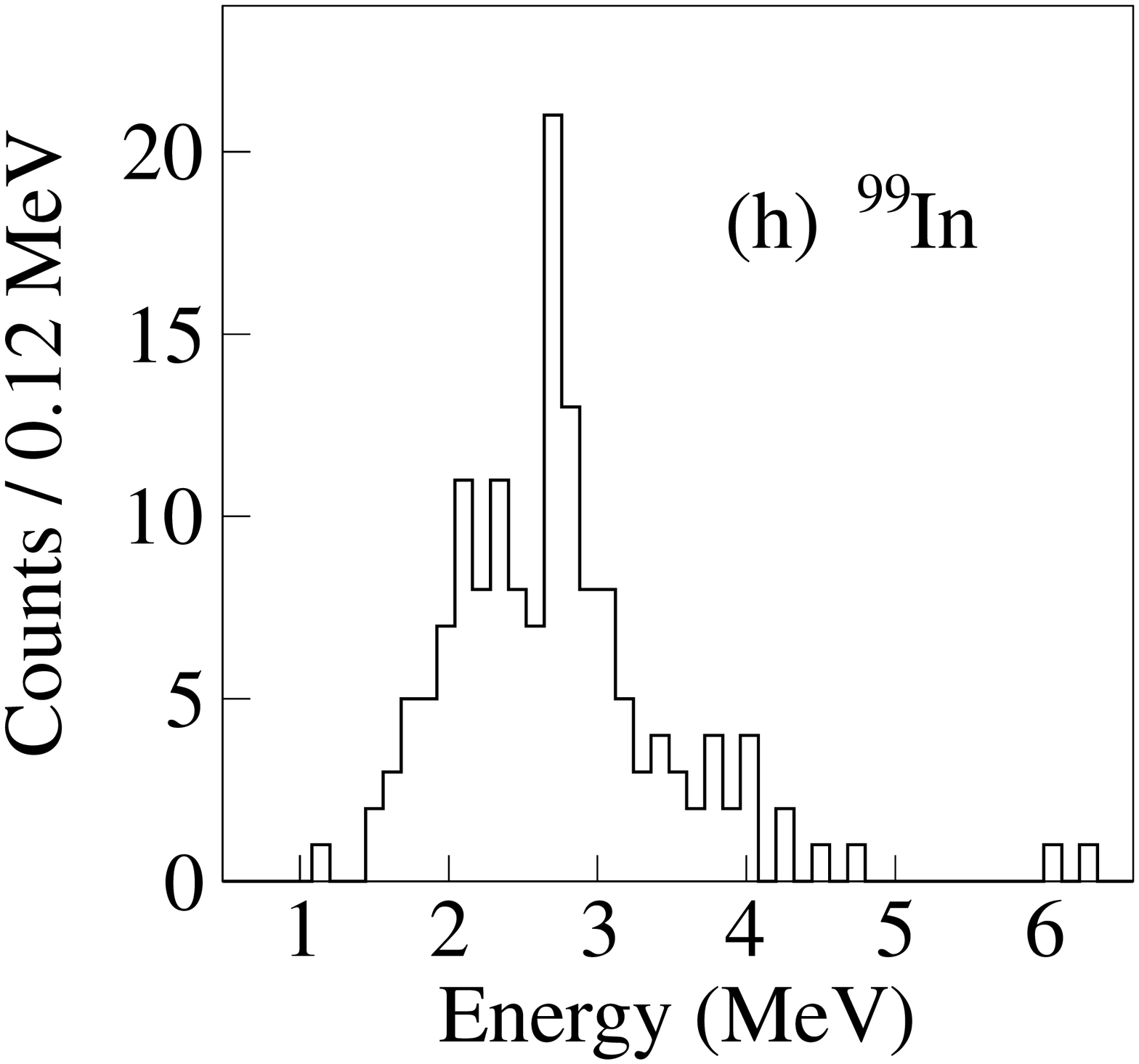}}
\hspace{2mm}
 \subfigure{\label{in98}\includegraphics[width=0.2187\textwidth]{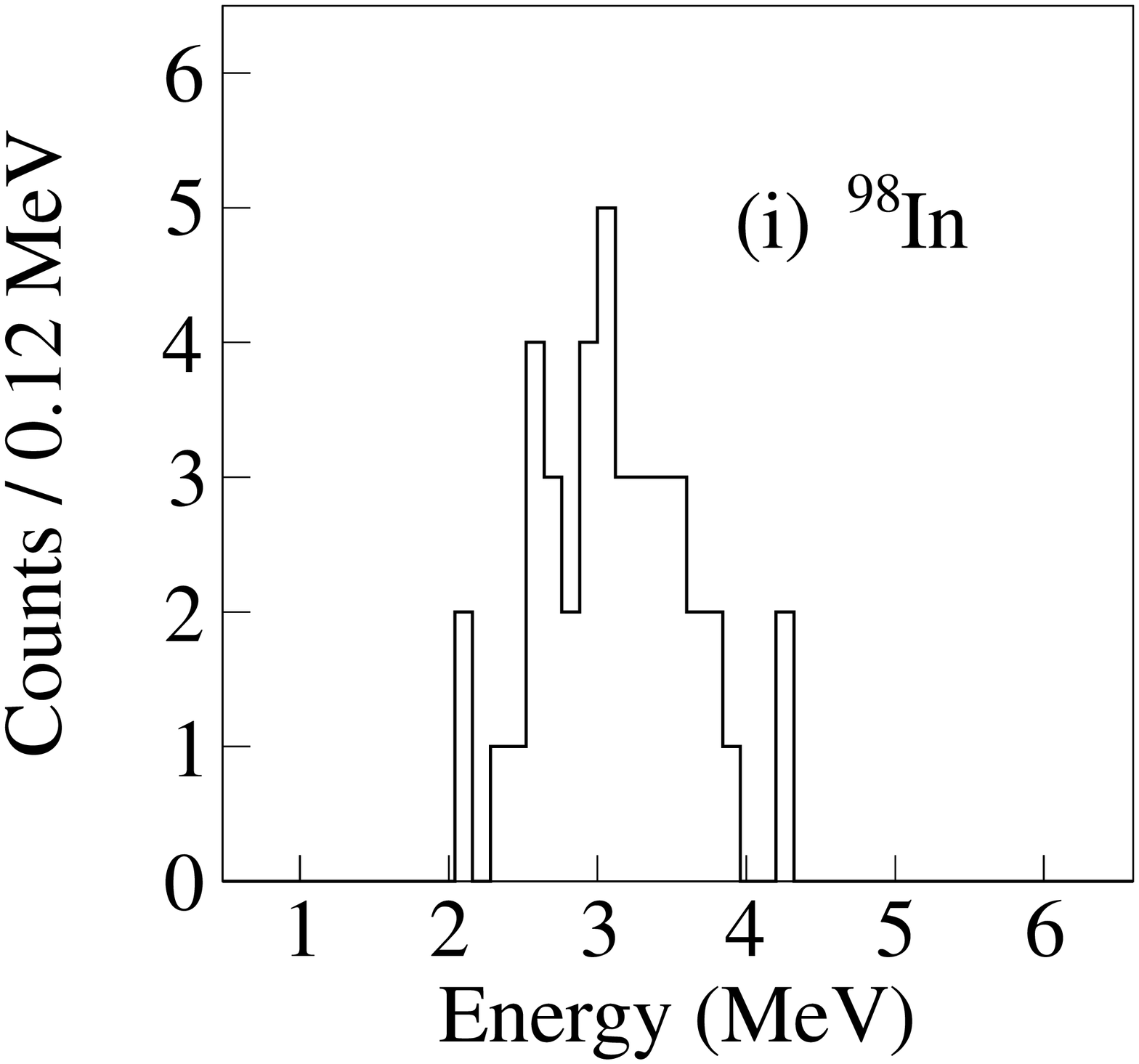}}
\hspace{2mm}
 \subfigure{\label{sn101}\includegraphics[width=0.2187\textwidth]{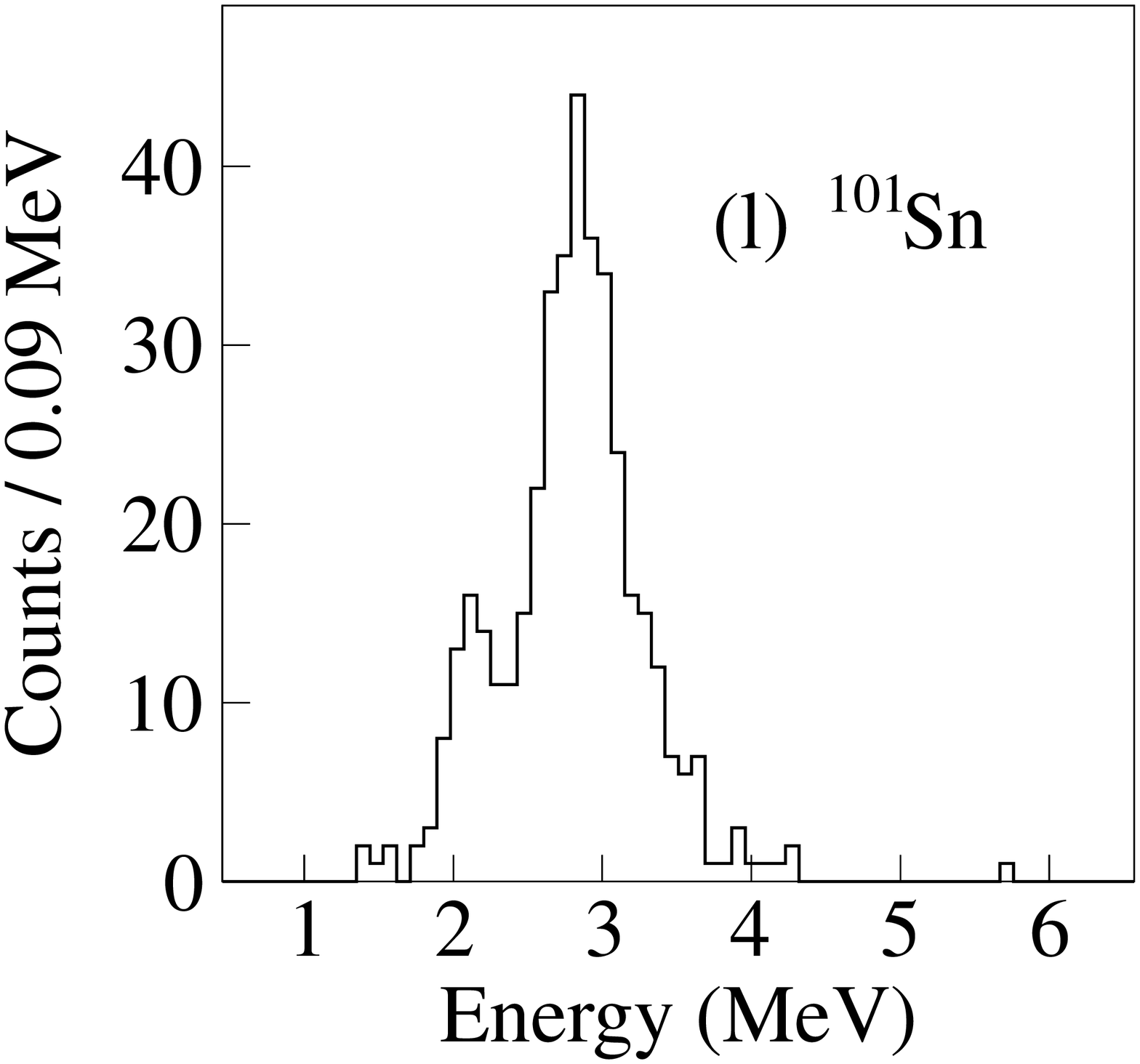}}
\end{center}
\vspace{-5.8mm}
\caption{Energy deposited in the DSSD by $\beta p$-emission events correlated with several implanted nuclides.}
\label{pspectra}
\end{figure}

\begin{figure}[h!]
\centering
\vspace{.6mm}
\hspace{-3.mm}
\includegraphics[width=0.49\textwidth]{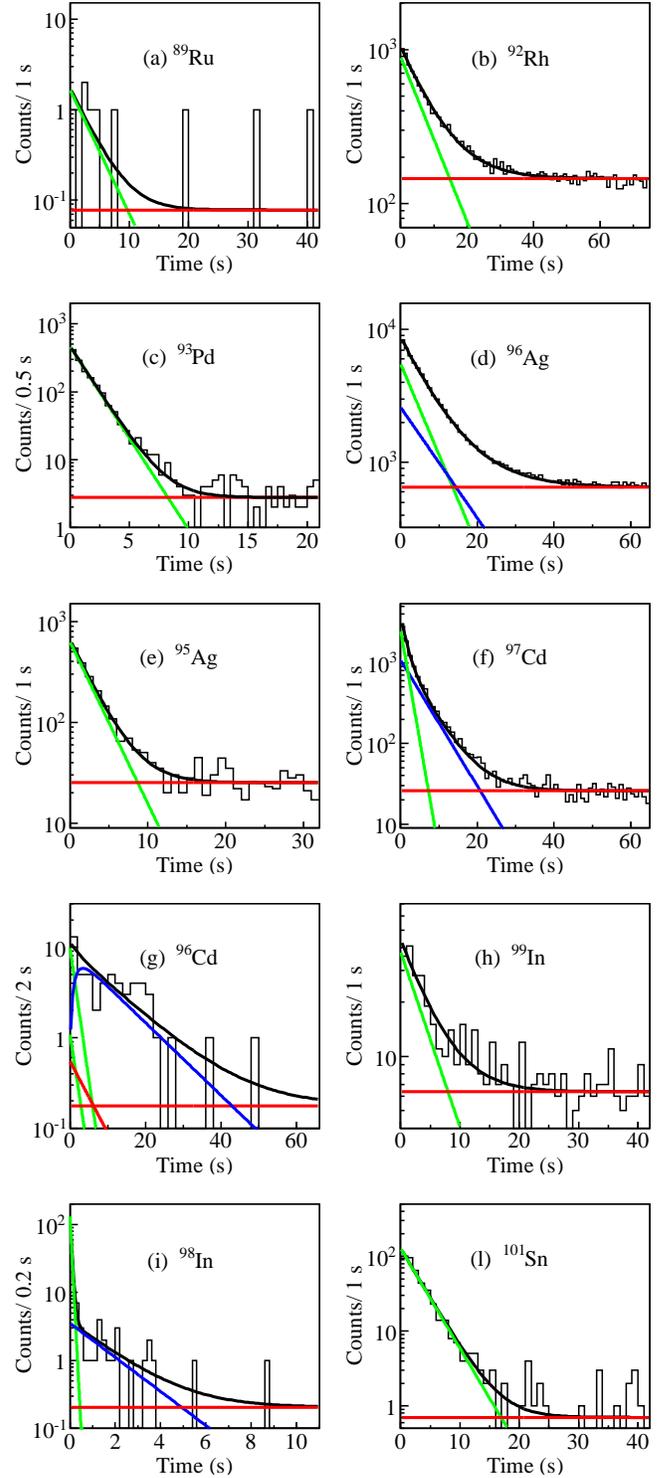}
\vspace{-5.8mm}

\caption{(color online). Time distribution of $\beta p$ events correlated with several implanted nuclides. Fits take into account the exponential $\beta p$ decay of parent ground (green) and isomeric states (blue), and a constant background (red). In the case of $^{96}$Cd, the contribution of the $\beta$-decay daughter $^{96}$Ag (blue), the $^{97}$Cd contamination (red) had to be included. The sum of all components is also shown (black).}
\label{ptime}
\end{figure}

In Fig.~\ref{pspectra} and Fig.~\ref{ptime} are presented the energy spectra and $\beta p$-decay curves , respectively, for several implanted nuclei. The deduced half-lives and $b_{\beta p}$ values are summarized in Table~\ref{tab_results}. The decay properties of these nuclides are discussed in detail in the following sections.

\subsection{The $\beta p$ precursor $^{89}$Ru}
The nucleus $^{89}$Ru was first identified at NSCL following fragmentation of a $^{92}$Mo beam \cite{Yen92}. The first $\beta$-decay study of this nucleus was conducted at Heavy Ion Research Facility in Lanzhou (HIRFL), where $^{89}$Ru nuclei were produced in the reaction $^{58}$Ni($^{36}$Ar, 2p3n) and were uniquely identified by tagging on coincidences between protons and known $\gamma$ rays from the deexcitation of levels in $^{88}$Mo \cite{Zha99}. A half-life of 1.2(2)~s was adopted for $^{89}$Ru based on coincidences with the 741~keV transition in the $\beta p$ daughter. These results were questioned in a second $\beta$-decay study of $^{89}$Ru \cite{Dea04}, where a reported $b_{\beta p}$$<$0.15\% was considered incompatible with the results reported in Ref.~\cite{Zha99}.

In the experiment reported here, a total of 167 $^{89}$Ru ions were implanted in the DSSD and six $\beta p$ events were identified. A $b_{\beta p}$ of 3$^{+1.9}_{-1.7}$\% was deduced, considering that one of the $\beta p$ events might be attributed to background, based on the measured background rate. The uncertainty was determined assuming a binomial probability distribution and applying a 95\% confidence level. A fit to the $\beta p$-decay curve in Fig.~5(a) resulted in a deduced half-life of 2.2(12)~s, consistent with the results in Ref.~\cite{Zha99}.    

\subsection{The $\beta p$ precursor $^{92}$Rh}

Results from $\beta$-delayed $\gamma$ spectroscopy studies of $^{92}$Rh were previously reported in Refs.~\cite{Zho99,Dea04}. In Ref.~\cite{Dea04}, the $\beta$-decay activity of $^{92}$Rh was attributed to two states, a 6$^{+}$ and a 2$^{+}$ state with half-lives of 4.66(25)~s and 0.53(37)~s, respectively. The energy difference between the two states was not determined. The presence of a 2$^{+}$ isomeric state may be significant for astrophysics, since the nucleus $^{92}$Rh lies along the rp-process path and is produced by proton capture on $^{91}$Ru \cite{Scha01}. The 2$^{+}$ state may be directly populated during the rp process due to its low spin (the $^{92}$Ru ground state spin is 0$^+$). However, no evidence for the 2$^{+}$ isomer was found in this work. The adopted half-life based on the analysis of the $\beta p$ and $\beta \gamma$ data was 5.63(7)~s, which is longer than both of the two previously reported half-lives. No evidence of multiple and disparate time components was found when the decay curve was gated on the 865.1~keV $\gamma$ peak that, according to Ref.~\cite{Dea04}, follows the $\beta$ decays of both the 6$^{+}$ and the 2$^{+}$ states. Considering the half-lives and the relative populations of the two states reported in Ref.~\cite{Dea04}, the measurement completed in this work should be sensitive to the existence of two time components in the $\beta$-decay curve of $^{89}$Ru. This discrepancy cannot be explained by a possible smaller production of the 2$^{+}$ state in the present experiment, as the relative intensities of $\beta$-delayed $\gamma$ rays observed in this work are compatible with Ref.~\cite{Dea04} (see Table~\ref{tabRh92}). In particular, the intensity ratio of $\gamma$ rays with energies 865.1~keV and 990.4~keV, which are fed by both states in the first case, and only by the 6$^{+}$ state in the second, would indicate a higher production rate of the 2$^{+}$ state in the experiment reported here.

Placement of the $\gamma$-ray transition with energy 919.1~keV in Ref.~\cite{Dea04} disagrees with the results of a previous in-beam experiment \cite{Zho99}, the coincidence data obtained here support the assignment made in Ref.~\cite{Dea04}.

\begin{table}[ht!]
 \caption{Energy and relative intensity of $\gamma$ rays following $\beta$ and $\beta p$ decay of $^{92}$Rh. Intensities are determined neglecting internal conversion. In the cases with sufficient statistics, the $\gamma$ gated half-life is reported.}
 \centering
 \vspace{8pt}
 \begin{tabular}{c | c c c }

    \hline
\vspace{-3mm}\\
&   E$_{\gamma}$  & ~~~~ Relative~~~~~& Half-life \\  
&    (keV)  & ~intensity (\%)& (s)\\  
\vspace{-3mm}\\
     \hline
&161.7(2)  &  12(1)       & \\
&339.4(7)  &  34(2)       & 3.90(80)\\
$\beta$ &817.2(5)  &  73(3)       & 5.85(25) \\
&865.1(3)  &  140(4)      & 5.35(20)  \\
 &918.4(1)  &  9(1)        &\\
&990.4(4)  &  100         & 6.58(25) \\
\hline
         &394.4(2)  & 100(20) &  4(1) \\
$\beta p$&893.1(2)  & 98(20)    &  5(1)\\
        &1097.5(5) & 15(10)  &\\
   \hline
 \end{tabular}
  \label{tabRh92}
  \end{table}

\begin{figure}[ht]
\centering
\includegraphics[angle=0,width=0.5\textwidth]{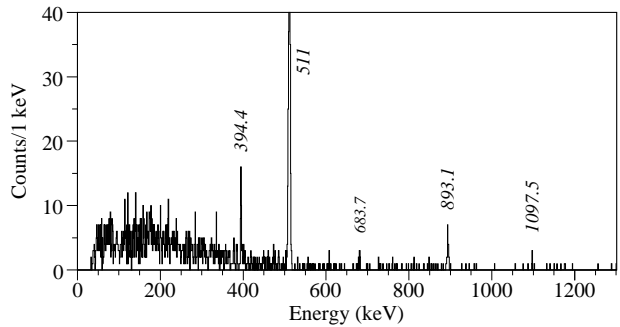}
\vspace{0pt}
\caption{$\gamma$-ray spectrum coincident with protons following the $\beta$ decay of $^{92}$Rh. The 683.7~keV $\gamma$ rays are emitted following the $\beta p$ decay of the contaminant $^{97}$Cd.}  
\label{rh92gamma}
\end{figure}

$\beta p$ emission from $^{92}$Rh is reported here for the first time [Fig.~\ref{rh92}]. A $b_{\beta p}$ of 1.9(1)\% was deduced, and a $\gamma$-ray spectrum coincident with $\beta p$ events is presented in Fig.~\ref{rh92gamma}. The $\gamma$ rays with energies 394.4(2), 893.1(2), and 1097.5(5)~keV originate from the deexcitation of known levels in $^{91}$Tc and have relative intensities of 97(20)\%, 100\%, and 14(9)\%, respectively. 

\subsection{The $\beta p$ precursor $^{93}$Pd}
$^{93}$Pd is predicted to be a waiting point in the astrophysical rp process~\cite{Scha98} and its identification was first reported in Ref.~\cite{Hen94}. A half-life value of $9.3_{-1.7}^{+2.5}$ s was reported in Ref.~\cite{Wef99}, but was later revised to 1.0(2)~s \cite{Kie01}. Subsequent experiments confirmed the shorter half-life \cite{Schm00,Xu01}. 
A half-life of 1.15(5)~s was deduced in this work based on a fit of the $\beta p$-decay curve, again in agreement with the most recent measurements in the literature.

\begin{figure}[ht]
  \includegraphics[width=0.5\textwidth]{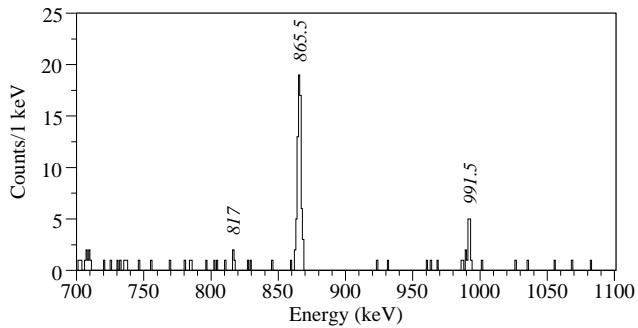}

 \caption{Section of the $\gamma$-ray spectrum measured in coincidence with $\beta$-delayed protons from the decay of $^{93}$Pd. The labeled peaks correspond to known deexcitations in the $\beta p$ daughter $^{92}$Ru 2$^+$$\rightarrow$0$^+$ at 865.5~keV and 4$^+$$\rightarrow$2$^+$ at 991.6~keV. Three counts at 817~keV suggest the weak population of the 6$^{+}$ state at 2671.5~keV in $^{92}$Ru and its subsequent decay into the lower 4$^+$ state.}
  \label{pd93psega}
\end{figure}

$\beta$-delayed proton activity from $^{93}$Pd was reported in Ref.~\cite{Schm00}, where an upper limit of $b_{\beta p}<5\%$ was observed. Ref.~\cite{Xu01} reported the detection of two $\gamma$ transitions with energies 865.7(5) and 991.5(10)~keV following the $\beta$-delayed proton emission of $^{93}$Pd. The intensities of the observed $\gamma$-ray transitions were compared with statistical model predictions, and found to be consistent with a $J^{\pi}=9/2^+$ ground state of $^{93}$Pd, although $J^{\pi}=7/2^+$ could not be excluded \cite{Xu01}. Decay of a 9/2$^+$ ground state of $^{93}$Pd is expected to weakly populate the 6${^+}$ state in $^{92}$Ru, resulting in the emission of a 817~keV $\gamma$ ray. While this $\gamma$-ray transition was not observed in the experiment reported in Ref.~\cite{Xu01} evidence for this specific decay pathway was found in this work (see Fig.~\ref{pd93psega}). The intensities of the three observed $\gamma$-ray transitions with energies 865.5(2), 991.5(3), and 817~keV are 100, 15.5(60)\%, and $<$8\%, respectively, and are compatible with the relative $\beta p$-branching ratios to different states in $^{92}$Ru predicted for the $\beta p$ decay of a 9/2$^+$ $^{93}$Pd ground state \cite{Xu01}.



The total $\beta$-delayed proton branch for $^{93}$Pd was deduced to be 7.5(5)\%, similar to the upper limit of 5\% reported in Ref.~\cite{Schm00}.
Taking the statistical model calculations \cite{Xu01} at face value, this new result is more consistent with a 7/2$^+$ ground state for $^{93}$Pd ($b_{\beta p} =4.8$\%) as opposed to the previously proposed 9/2$^+$ ground state ($b_{\beta p} =1.7$\%). The energy spectrum of $\beta$-delayed protons, shown in Fig.~4(b), agrees with that in Ref.~\cite{Schm00}, confirming the absence of significant isobaric contamination in the latter. 

A question regarding the decay of $^{93}$Pd is the assignment of the $\beta$-delayed $\gamma$ ray transition with energy 864.1~keV, initially reported in Ref.~\cite{Schm00}. This $\gamma$-ray transition was assigned to the decay of a predicted 13/2$^+$ state in the daughter $^{93}$Ru. An alternative assignment was made in Ref.~\cite{Xu01}, where $\gamma$ rays of similar energy were observed in coincidence with $\beta$-delayed protons, and assigned to the deexcitation of the first 2$^+$ state in the $\beta p$-daughter $^{92}$Ru.

A peak in the $\gamma$-ray spectrum at energy 865.5~keV was seen in both the $\beta$- and $\beta p$-delayed $\gamma$ ray spectra reported here, but the decay time of this $\gamma$-ray transition gated on detected $\beta$ particles was not consistent with the known half-life of $^{93}$Pd. Therefore, the 865.5~keV transition observed following the decay of $^{93}$Pd is predominantly produced by $^{92}$Ru, as suggested in Ref.~\cite{Xu01}.

\subsection{The $\beta p$ precursor $^{96}$Ag}
\label{Ag96}

The nucleus $^{96}$Ag is known to have two low-energy $\beta$-decaying states, with tentatively-assigned spins and parities of 8$^{+}$ and 2$^{+}$ \cite{Bat03}. A half-life of 4.39(8)~s was deduced for the decay of the shorter-lived, presumed 8$^+$ state by gating the decay curve of $^{96}$Ag by the 325~keV $\gamma$-ray transition in $^{96}$Pd. A half-life of 6.8(10)~s was deduced for the longer-lived, proposed 2$^+$ state via a two-component fit of the $\beta p$ activity, where the short-lived component of the decay curve was fixed to a half-life value of 4.39~s [Fig.~5(d)]. The same fit provided the relative contributions of the two $^{96}$Ag states to the $\beta p$ activity. The relative contributions of the two $\beta$-decaying states to the overall $\beta$ activity were obtained by comparing the intensities of the 325~keV and 1415~keV $\gamma$ rays. The 325~keV $\gamma$ ray is mainly fed by the $\beta$ decay of the presumed 8$^{+}$ state, while the 1415~keV is fed by both states. 
The population of the short- and long-lived states was determined to be 78(10)\% and 22(10)\%, with $b_{\beta p}=6.50(80)\%$ and 14(3)\%, respectively. The half-lives and $b_{\beta p}$ values agreed with those reported in Ref.~\cite{Bat03}. 

\begin{figure}[h]
  \centering
  \includegraphics[width=0.52\textwidth]{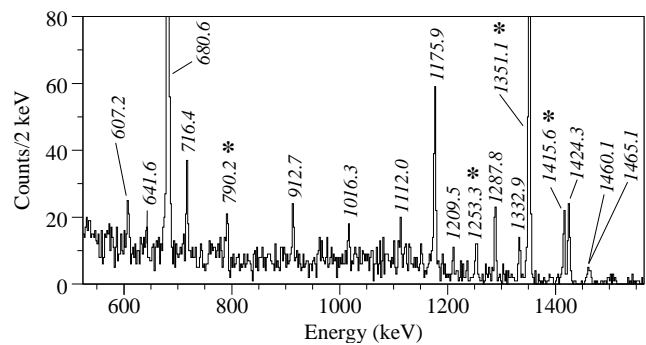}
  \caption{Section of the measured $\gamma$ ray spectrum in coincidence with $\beta p$ decays of $^{96}$Ag. The transitions attributed to the deexcitation of states in $^{95}$Rh are labeled with the corresponding energy. Asterisks mark transitions due to the $\beta p$ decay of the contaminant $^{97}$Cd.}
  \label{Rh95}
\end{figure}

The excitation energy of the $^{95}$Rh states reported in Ref.~\cite{Bat03} with energies 1180, 1430, 1570, and 2080~keV were determined with much higher accuracy using the $\beta p$-delayed $\gamma$ spectrum shown in Fig.~\ref{Rh95}. The $\gamma$-ray transition with energy 913~keV, previously reported from in-beam spectroscopy of $^{95}$Rh and attributed to the deexcitation of a 2264~keV state \cite{Rot94}, was also observed.
In addition, several weak $\gamma$ transitions not reported previously, were observed here that likely originate from the deexcitation of states in $^{95}$Rh but could not be placed in the $^{95}$Rh level scheme. All transitions observed from the deexcitation of $^{95}$Rh states are listed in Table~\ref{rh95tab}.

\begin{table}[ht]
  \caption{ $^{96}$Ag $\beta p$-delayed $\gamma$ rays assigned to transitions in $^{95}$Rh with half-lives and relative intensities. Intensities are not corrected for internal conversion.}
  \centering
  \vspace{8pt}
  \begin{tabular}{c  c  c   }
    \hline
\vspace{-6mm}\\
&&\\
  Peak energy & ~~half-life & ~~~I~~~\\  
    (keV)     &   ~~ (s)      &     ~~(\%)      \\
\vspace{-6mm}\\
&&\\
    \hline
\vspace{-5mm}\\
&&\\
    607.2(4)   &~~5.1(11)  &      ~~~  6(1)      \\
    641.6(6)   &         &      ~~~ 3(1)        \\
    680.6(1)   &~~6.4(4)   &      ~~~ 100\\
    716.4(5)    &~~4.6(10)  &     ~~~7(1)  \\
    912.7(5)    &~~4.8(10)  &  ~~~7(1)    \\
    1016.3(4)     &         &  ~~~5(1) \\
    1112.0(2)    &          &~~~4(1)\\
    1175.9(2)   &~~5.1(8)     &~~~ 24(3)    \\
    1209.5(6)   &           &~~~3(1)\\
    1287.8(2)   &~~6.9(4)     & ~~~10(2)    \\
    1332.9(4)   &~~6.2(18)   & ~~~6(1)      \\
    1351.1(1)   &~~4.25(4)   & ~~~75(6) \\
    1424.3(2)   &~~3.9(8)   & ~~~11(2)  \\
    1460.1(5)   &       &  ~~~2(1)\\        
    1465.1(6)        &      & ~~~2(1)           \\
 \vspace{-2mm} \\
   \hline
  \end{tabular}
  \label{rh95tab}
  \end{table}

\subsection{The $\beta p$ precursor $^{95}$Ag}

The first identification of $^{95}$Ag was reported in Ref.~\cite{Schm94}, along with a half-life value of 2.0(1)~s deduced from $\beta p$-decay data. An improved half-life of 1.74(13)~s was reported later, based on $\beta\gamma$-coincidence data \cite{Schm97}. The discrepancy between the two half-lives was explained by the presence of $^{95}$Pd contamination in the earlier $\beta p$ study. Long-lived 1/2$^-$ and 23/2$^-$ isomers, predicted for $^{95}$Ag \cite{Oga83}, were observed to decay predominantly by $\gamma$ emission with short lifetime limits of $<$500~ms and $<$16~ms, respectively \cite{Dor03,Mar03}. The $\beta$ decay of $^{95}$Ag is therefore likely dominated by the decay of the ground state. 

The $\beta p$-energy spectrum [Fig.~\ref{ag95}] and the $\beta p$-decay half-life of 1.80(8)~s obtained in the present work [Fig.~5(e)] agree with the results reported in Ref.~\cite{Schm97}, resolving the discrepancy with the earlier $\beta p$-decay data Ref.~\cite{Schm94} and supporting the conclusions of Ref.~\cite{Schm97}. A $b_{\beta p}$ value of 2.5(3)\% was also measured. No evidence was found here for $\beta p$ decay of the 21/2$^+$ isomeric state in $^{95}$Pd, which might be populated following the $\beta$ decay of $^{95}$Ag. The high-spin isomer in $^{95}$Pd has a 13~s half-life and a $b_{\beta p}$ branching of 0.74(0.19)\%. 

\begin{figure}[h]
\centering
\includegraphics[angle=0,width=0.52\textwidth]{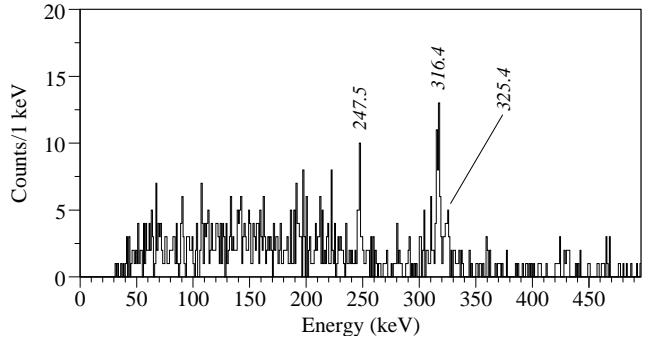}
\vspace{0pt}
\caption{Section of the $\gamma$ spectrum coincident with $\beta$-delayed protons from the decay of $^{95}$Ag. The $\gamma$ ray at 325.4~keV is a contamination from the $\beta p$ decay of $^{97}$Cd.}
\label{Ag95gammap}
\end{figure}

Two $\gamma$ rays with energies 247.5(20) and 316.4(3)~keV were observed in coincidence with $\beta p$ decay events from $^{95}$Ag (see Fig.~\ref{Ag95gammap}). Their half-lives, 1.37(40)~s and 1.70(25)~s, respectively, are compatible with the known half-life of $^{95}$Ag, but transitions with these energies are not known in $^{94}$Rh. 

The fit of the $\beta$-decay curve of $^{95}$Ag was also used to the determine the half-life of $^{95}$Pd. The fit was performed for $\beta$ decays detected more than 3~s after the implantation of $^{95}$Ag ions to eliminate the early-time contribution of conversion electrons from the $\gamma$ decay of $^{95}$Ag \si{\milli\second} isomers. The half-life of the granddaughter $^{95}$Rh was fixed to 5.02(10)~s \cite{Wei75} since the decay of $^{95}$Pd is not expected to populate the 1/2$^-$ isomer in $^{95}$Rh, as in nuclei with similar structure to $^{95}$Pd such as $^{93}$Ru and $^{91}$Mo. This decay curve analysis yielded in half-lives of 8.0(6) and 1.81(13)~s for $^{95}$Pd and $^{95}$Ag, respectively.

\subsection{The $\beta p$ precursor $^{97}$Cd}

The nucleus $^{97}$Cd was first identified at the ISOLDE on-line isotope separator through the observation of $\beta p$ emission \cite{Elm78}. Almost 20 years later, the experiment was repeated, where a $\beta p$-energy spectrum was obtained and a half-life of 2.8(6)~s was deduced for the ground-state decay \cite{Schm97}. However, shell-model calculations predict two isomeric states, a 1/2$^{-}$ state below 1~MeV, and a 25/2$^{+}$ state at about 2.4~MeV in $^{97}$Cd \cite{Oga83,Schm97}.
Both states are predicted to have sizable $\beta$-decay branches, and given the expected large $Q_{EC}-S_p$ window, both states should exhibit $\beta p$ emission. 

The results for the decay of $^{97}$Cd were previously published \cite{Lor11}, and a summary is provided here for completenes. Evidence for high-spin isomer was found, based on the different half-lives of $^{97}$Cd $\beta p$-decay events coincident with $\gamma$ rays from the deexcitation of states in $^{96}$Pd. In addition, two components were evident in the $^{97}$Cd decay curve as shown in Fig.~5(f). Half-life values of 1.10(8) and 3.8(2)~s, and $b_{\beta p}$ values of 12(2)\% and 25(4)\% were deduced for ground and isomeric states, respectively. Details can be found in Ref.~\cite{Lor11} and Table~\ref{tabCd97}. 

In Fig.~10(a) is given the energy spectra of $\beta p$ events from the decay of the $^{97}$Cd ground state and of the 25/2$^+$ isomer.
The proton energy spectrum for the 25/2$^{+}$ isomeric state was obtained from the activity observed between 5 and 10~s following the implantation of $^{97}$Cd. Within this time window, no significant contribution from the 1.1~s component due to the $^{97}$Cd ground state decay is expected.
The isomeric spectrum was subtracted from the $\beta p$-event spectrum during the first 1~s after an implantation, to obtain the $\beta p$ spectrum following decay of the $^{97}$Cd ground state The isomeric contribution was derived from the shape measured at late times, scaled by the deduced half-life and the population of the isomeric state.
The $\beta p$-event spectrum from the ground state decay includes a second peak at $\sim$4.5 MeV, which is reproduced by shell-model calculations [Fig.~10(b)] \cite{Schm97}.


\begin{table}[ht]
  \caption{Energy, relative intensity, and half-lives of $^{97}$Cd $\beta p$- and $\beta$-delayed $\gamma$ ray transitions. Intensities are corrected by internal conversion $IC$.}
  \centering
  \vspace{8pt}
  \begin{tabular}{c  c  c  r l c}
    \hline
\vspace{-2mm}
                      &          &             &                              &    \\
    Energy of $^{96}$Pd & J$^{\pi}$ & E$_{\gamma}$ &\multicolumn{2}{c}{Half-life}& I$_{\beta p} $ \\  
    level (keV)         &           &   (keV)     &  &~~(s) & (\%)     \\
\vspace{-3mm}
                      &          &             & \\
    \hline
    4574.9(3)             & 12$^{+}$ &  790.2(2) &   &3.5(4)    &17(2)   \\
    3783.7(2)              & 10$^{+}$ & 1253.3(1) &   &4.0(3)   &58(5)  \\
    2531.4(1)              & 8$^{+}$  &  106.6(9) &   &3.6(3)   &63(5)\tmark[a]   \\
    2424.8(1)              & 6$^{+}$  &  325.4(6) &   &3.5(2)   &67(4)\tmark[b]   \\
    2099.4(1)              & 4$^{+}$  &  684.9(1) &   &3.65(25) &77(5)    \\
    1415.5(8)              & 2$^{+}$  & 1415.5(1) &   &2.29(15) &100   \\
    \hline
\vspace{-2mm}
                      &          &             &                              &    \\
    Energy of $^{97}$Ag &   &     &   & & I$_{\beta} $ \\  
    level (keV)         &           &    &   & & (\%)     \\   
\vspace{-3mm}
                      &          &             & \\
    \hline
    6225.6(5)                &   (27/2$^{+}$) & 1305.6(4)        &  &           & 16(4)    \\
    4920.0(4)                &   (23/2$^{+}$) & 2575.3(3)        &  &           & 22(5)     \\
    2344.7(2)                &   (21/2$^{+}$) & 291.3(2)         & & 3.75(60)   & 77(10) \tmark[c]     \\
    2053.4(2)                &   (17/2$^{+}$) & 763.1(1)         & & 3.8(6)     & 119(12)\\
    1290.3(1)                &   (13/2$^{+}$) & 1290.3(1)        & & 3.85(70)   & 100\\
    716.2(2)                 &   (7/2$^{+}$)  & 716.2(2)         & & 1.1(2)         & 40(5)\\  
    \hline
 \end{tabular}
\flushleft
\hspace{3mm}

  \label{tabCd97}
  \end{table}

\begin{figure}[h]
  \centering
  \subfigure{\label{cd97schmexp}\includegraphics[width=0.23\textwidth]{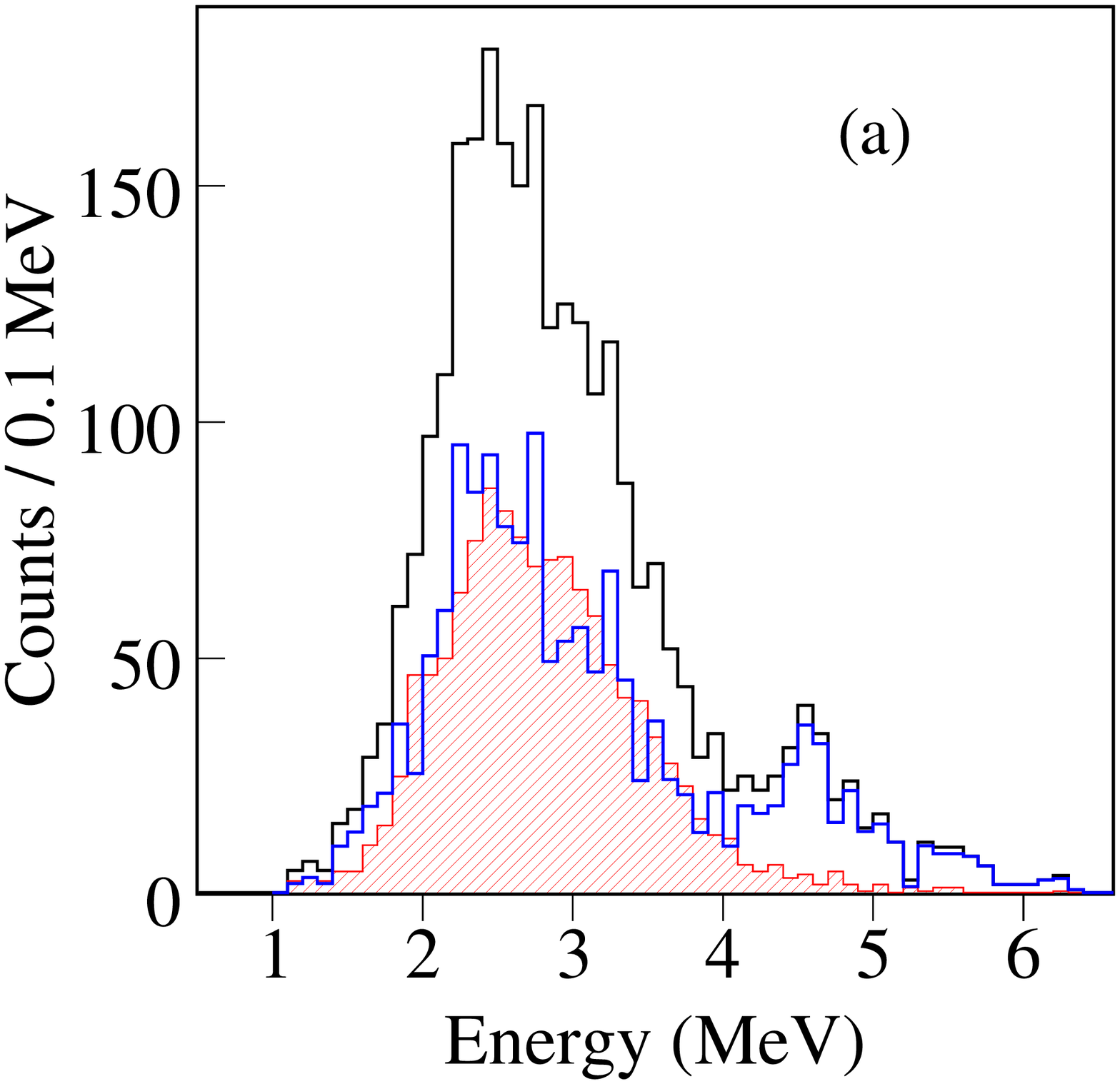}}
  \subfigure{\label{cd97schmth}\includegraphics[width=0.23\textwidth]{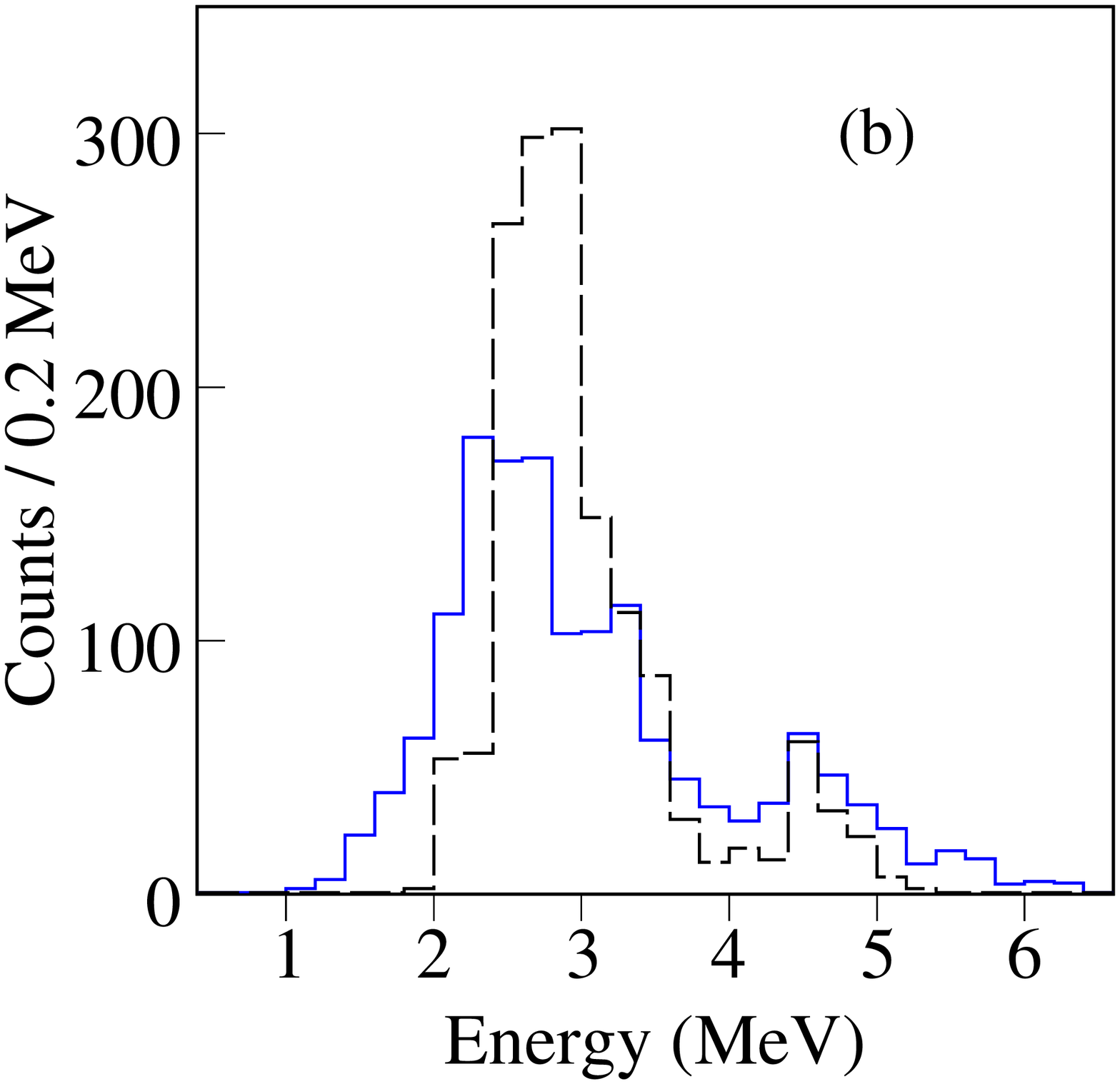}}
  \vspace{-2pt}
  \caption{(color online). (a) $\beta p$-energy spectrum of all $^{97}$Cd decays (black), of 25/2$^{+}$ $^{97}$Cd isomeric decays (filled red), and $^{97}$Cd ground-state decays (blue). (b) Comparison between the $\beta p$-energy spectrum of the $^{97}$Cd ground-state decays deduced from this experiment (blue) and predicted by shell-model calculation \cite{Schm97} (dashed black).}
  \label{cd97schmth}
\end{figure}

\subsection{The $\beta p$ precursor $^{96}$Cd}
\label{Cd96}
$^{96}$Cd is a possible waiting point in the astrophysical rp process and an rp-process progenitor of the p nucleus $^{96}$Ru \cite{Baz08, Scha98}. The first half-life measurement of 1.03$^{+0.24}_{-0.21}$~s, reported in Ref.~\cite{Baz08}, was later revised to 0.67(15)~s \cite{Sin11}. Additional details for this nucleus are given below. 

Unlike the other isotopes studied in this work, analysis of the $\beta p$-decay data of $^{96}$Cd was complicated by feeding pathways outside the simple parent decay. First, $\beta p$ emission from the $\beta$-daughter $^{96}$Ag was not negligible, and this contribution was included in the analysis with $b_{\beta p}$ and the half-life of $^{96}$Ag considered as free parameters. Second, contamination of $^{97}$Cd in the events identified as $^{96}$Cd implantations needed to be considered. The $^{97}$Cd contribution was estimated using $b_{\beta p}$ values and relative populations of the two $\beta p$-decaying $^{97}$Cd states deduced in this work. The half-life of $^{96}$Cd was fixed to the value deduced in this experiment based on $\beta$-decay data \cite{Baz08}. Finally, a constant background was included, determined from the decay curve data at later times out to 200~s. An initial fitting of the $^{96}$Cd $\beta p$-decay curve was performed with four components and three free parameters: (i) $b_{\beta p}$ of $^{96}$Cd, (ii) $b_{\beta p}$ of $^{96}$Ag, and (iii) the half-life of $^{96}$Ag. A half-life value of 7.5(15)~s was obtained for $^{96}$Ag, indicating that the dominant $\beta$-decay branching of $^{96}$Cd feeds the 2$^+$ state in $^{96}$Ag. The decay curve fitting was then repeated with the $b_{\beta p}$ and half-life value of $^{96}$Ag fixed to the known values for the 2$^{+}$ state. The resulting fit of the decay curve is shown in Fig.~5(g).

The analysis was performed under several assumptions for the level of $^{97}$Cd contamination in the $^{96}$Cd particle identification gate. The consideration of implantations with at least 10\% (241 events), 50\% (189 events), and 90\% (112 events) probability to be $^{96}$Cd resulted in consistent $b_{\beta p}$ values confirming the correct treatment of the $^{97}$Cd contamination. The result of $b_{\beta p}=$ 5.5(40)\% was obtained for events with at least 50\% probability to be $^{96}$Cd implantations. The reported error in the $b_{\beta p}$ value is largely attributed to the uncertainties of the half-life of $^{96}$Cd and the $b_{\beta p}$ of the 2$^{+}$ state in $^{96}$Ag. 

No evidence for the presence of a $\beta p$ decay of the 16$^+$ isomeric state \cite{Sin11} could be found, which is not surprising given the limited data collected for the $^{96}$Cd decay.

\subsection{The $\beta p$ precursor $^{98}$In}

\begin{figure}[h!]
\centering
  \vspace{-1mm}
\subfigure{\includegraphics[width=0.348\textwidth]{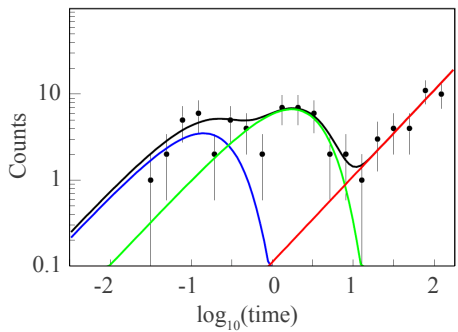}}
\caption{(color online). Decay curve and fit of the $\beta p$ activity of $^{98}$In in logarithmic-time binning. The fit includes the contribution of the ground state of $^{98}$In (blue), the isomeric state of $^{98}$In (green), and background (red). The sum of all components is also displayed (black).}
\label{in98timelog}
\end{figure}

The first experiment to produce $^{98}$In \cite{Kie01} revealed evidence for two $\beta$-decaying states with half-lives 32$^{+32}_{-11}$~ms and 1.2$^{+1.2}_{-0.4}$~s, respectively \cite{Kie01}. The shorter component was attributed to the superallowed Fermi decay of the proposed 0${^+}$ ground state. Analysis of the $\beta$ decay of this nucleus in the present experiment confirmed the presence of two time components in the decay curve data. Half-life values of 47(13)~ms and 0.66(40)~s deduced from a fit to the decay curve \cite{Baz08} are consistent with the results reported in Ref.~\cite{Kie01}.
The analysis of $\beta$-delayed proton emission from $^{98}$In is reported here for the first time. The time distribution of 63 $\beta p$ events again showed evidence of two time components [Fig.~5(i) and Fig.~\ref{in98timelog}]. Subsequent decay curve fitting produced half-lives of 135(65)~ms and 1.27(30)~s, and branchings of $b_{\beta p}$=5.5${^{+3}_{-2}}$\% and 19.5(13)\%, for the ground state and isomeric state, respectively. The deduced $\beta p$ half-life is more precise for isomeric state than that reported in Ref.~\cite{Baz08}. The higher precision is attributed to a reduction of background and elimination of daughter-decay contributions that outweigh the reduced statistics.
The observation of $\beta p$ activity stemming from the ground state of $^{98}$In is surprising, as the superallowed Fermi decay of the 0$^+$ $^{98}$In ground state is expected to directly populate the 0$^{+}$ ground state in $^{98}$Cd.

\subsection{The $\beta p$ precursor $^{101}$Sn}

Having just one additional neutron compared to the presumed doubly-magic nuclide $^{100}$Sn, $^{101}$Sn is of particular interest for probing the single particle level structure in this region. Recently, a $\gamma$ transition with energy 172~keV was identified in a in-beam $\gamma$-ray experiment by tagging on $\beta$-delayed protons from $^{101}$Sn. The single $\gamma$ ray was assigned to the deexcitation of the first excited 7/2$^+$ state to the 5/2$^+$ ground state, representing the g$_{7/2}$d$_{5/2}$ single particle level spacing \cite{Sew07}. A $\gamma$-ray transition of similar energy was observed in the $\alpha$-decay measurements of the chain $^{109}$Xe$\rightarrow ^{105}$Te$\rightarrow ^{101}$Sn, where an opposite level ordering was proposed with a 5/2$^+$ excited state and a 7/2$^+$ ground state \cite{Dar10}.

Shell-model calculations, making use of different effective interactions, account both orderings \cite{Sew07}. This is not surprising due to the small energy separation between the ground state and first excited state. However, it was shown in Ref.~\cite{Kav07} that the $\beta p$-emission properties of $^{101}$Sn, as predicted by the shell model, depend significantly on the spin of $^{101}$Sn, offering a model dependent way to discriminate between the two possibilities. The assumption of $J^\pi = 5/2^+$ or $7/2^+$ for the $^{101}$Sn ground state yields $b_{\beta p}=26\%$ or 14\%, respectively. The calculated proton-energy spectrum is expected to have a broad, single peak for the case of a 5/2$^{+}$ ground state for $^{101}$Sn while a highly structured spectrum with a second peak around 3~MeV is predicted for a 7/2$^{+}$ ground state for this nuclide. The calculations indicate this feature is robust with respect to the uncertain proton-separation energy of $^{101}$In, though the expected $\beta p$ branchings show a $Q$-value dependence.

A total of 400 $\beta p$ events were detected here for $^{101}$Sn. A half-life value of 2.1(3)~s was deduced based on the $\beta p$-decay curve in agreement with the weighted average of 1.7(3)~s from previous experiments \cite{Sew07,Kav07}. The $b_{\beta p}$ = 20(1)\% deduced for the $^{101}$Sn ground state decay agrees with a previous measurement ($b_{\beta p}$ = 14${^{+15}_{-10}}$\% \cite{Sto01}). This new $b_{\beta p}$ falls between the shell model predictions for the two possible ground state spin assignments for $^{101}$Sn. This is not surprising, since it was discussed in Ref.~\cite{Kav07} that the variation in proton-separation energy for $^{101}$In within its uncertainty, will give $b_{\beta p}$ values between 8\% and 48\% for $J^\pi = 5/2^+$ and between 2\% and 38\% for $J^\pi = 7/2^+$. More precise masses are needed to better constrain the shell-model calculations. 

The $\beta p$-energy spectrum [Fig.~4(l)] has a structure similar to that reported in Ref.~\cite{Sto01}, but has considerably more statistics. Nevertheless, it was not possible to draw a strong conclusion concerning the expectation of the statistical $\beta p$ decay for the two alternative spins within the shell-model framework. While the experimental spectrum in Fig.~4(l) shows little structure, in line with the prediction for a 5/2$^+$ ground state, there are events above 3~MeV that may be considered as a separate high-energy proton group that is expected to be present for the 7/2$^+$ decay based again on the shell model results \cite{Kav07}.

\begin{figure}
\centering
\includegraphics[angle=0,width=0.5\textwidth]{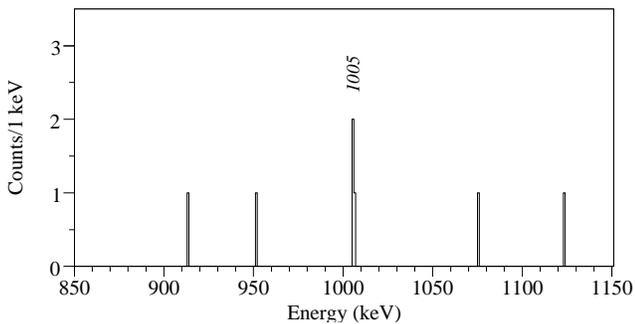}
\vspace{0pt}
\caption{Section of the $\gamma$ spectrum measured in coincidence with $\beta$-delayed protons from the decay of $^{101}$Sn. The labeled peak corresponds to the known deexcitation 2$^+$$\rightarrow$0$^+$ in the $\beta p$ daughter $^{100}$Cd.}
\label{sn101segap}
\end{figure}

No $\gamma$ rays following the $\beta$ decay of $^{101}$Sn were evident, consistent with shell-model results \cite{Kar06}, which predict fragmentation of the Gamow-Teller strength over many excited states in $^{101}$In. Four counts at energy 1004~keV were observed in coincidence with $^{101}$Sn proton events (Fig.~\ref{sn101segap}). This peak is likely from the deexcitation of the known first 2$^{+}$ state in $^{100}$Cd, fed by the $\beta p$ decay of $^{101}$Sn. No evidence for the 794~keV 4$^{+} \rightarrow 2^{+}$ transition in $^{100}$Cd was found, suggesting that the decay mainly feeds lower-spin states in the $\beta p$ daughter.

\subsection{The $\beta p$ precursor $^{100}$Sn}
\label{Sn100}
The production and separation of $^{100}$Sn was previously achieved in experiments both at GSI and at GANIL. At GSI, a 1 GeV/nucleon $^{124}$Xe beam was used to produce 7 $^{100}$Sn nuclei \cite{Sum97} and resulted in a half-life of 0.94$^{+0.54}_{-0.26}$~s. The data collected at GSI included the $^{100}$Sn implantations, and a half-life of 0.94$^{+0.54}_{-0.26}$~s. The GANIL experiment made use of a 60 MeV/nucleon $^{112}$Sn beam to produce 11 $^{100}$Sn \cite{Lew95}. An experiment employing the fusion-evaporation technique was also successful at GANIL and provided the mass of $^{100}$Sn \cite{Cha96}.

In this experiment, 14 $^{100}$Sn nuclei were implanted into the DSSD and a half-life of 0.55$^{+0.70}_{-0.31}$~s was deduced based on $\beta$-fragment correlation and the likelihood fit function \cite{Baz08}. The study of $\beta p$ emission from $^{100}$Sn was complicated by the contamination of $^{101}$Sn, which itself is a known $\beta p$ emitter. An attempt to estimate the $b_{\beta p}$ value for $^{100}$Sn was made by selecting 7 $^{100}$Sn implantations that, according to the probability-weighted particle identification from Ref.~\cite{Baz08}, had a contamination level of $^{101}$Sn below 5\%. No evidence for $\beta$-delayed protons correlated to any of the 7 $^{100}$Sn nuclei was found within 5~s of the $^{100}$Sn implantations. At the 95\% confidence limit, this result suggests that $b_{\beta p}<$35\% for the ground state of $^{100}$Sn.

\section{rp-process calculations}

\begin{figure}[hb]
\hspace{-20mm}
\includegraphics[angle=0,width=0.3\textwidth]{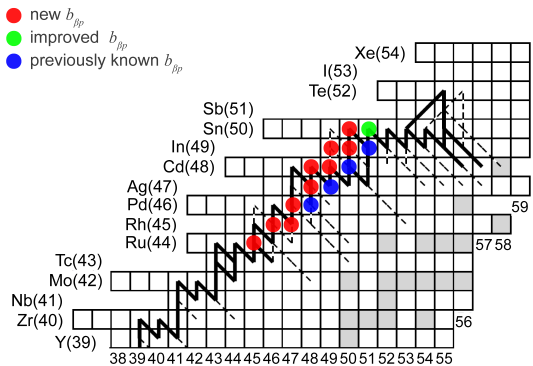}
\vspace{10pt}
\caption{(color online). Progenitors for $\beta$-delayed proton emission discussed in this paper. The solid and dashed lines denote the path of the rp process with reaction flows of more than 10\% and more than 1\% of the maximum flow, respectively. First measurements of $b_{\beta p}$ from this work are marked in red, improved measurements in green, and previously known values in blue.}
\label{chart}
\end{figure}

The implications of the new data for neutron-deficient nuclei below $^{100}$Sn on rp-process network calculations were explored using a single zone X-ray burst model \cite{Scha01}. The ReaclibV1 rates provided by the JINA Reaclib online database \cite{Cyb10} were used, as well as the new $b_{\beta p}$ and half-lives values. Burst ignition conditions were chosen to reflect an accretion rate of 10\% of the Eddington rate (0.1 $\dot{m}_{\rm Edd}$), a thermal flux out of the neutron star crust of 0.15 MeV per accreted nucleon, and a metallicity $Z= 10^{-3}$ of the accreted material. An accreted layer with solar metallicity would produce a similar burst at a higher accretion rate of 0.3 $\dot{m}_{\rm Edd}$. These conditions result in an extended rp process up to Te, and are therefore suitable to explore the general features of an rp-process mass flow in the Cd-Sn region (see Fig.~\ref{chart}). 

\begin{figure}
 \centering
\hspace{-1.mm} 
\subfigure{\label{bpresultsa}\includegraphics[width=0.45\textwidth]{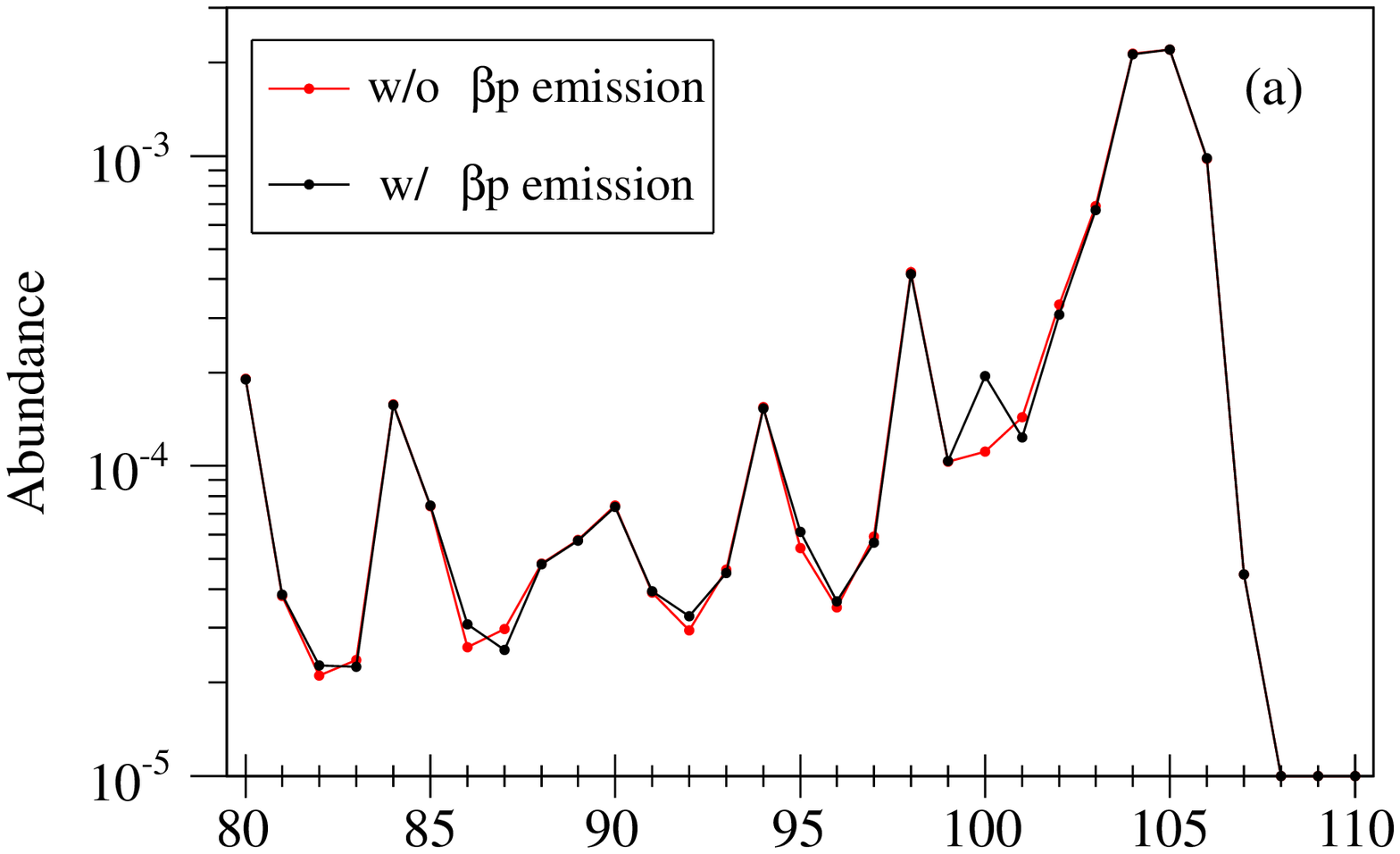}}
\centerline{\fontsize{5}{5}{Mass number}}
\vspace{-8mm}\\
\hspace{0.2mm}
\subfigure{\label{bpresultsb}\includegraphics[width=0.45\textwidth]{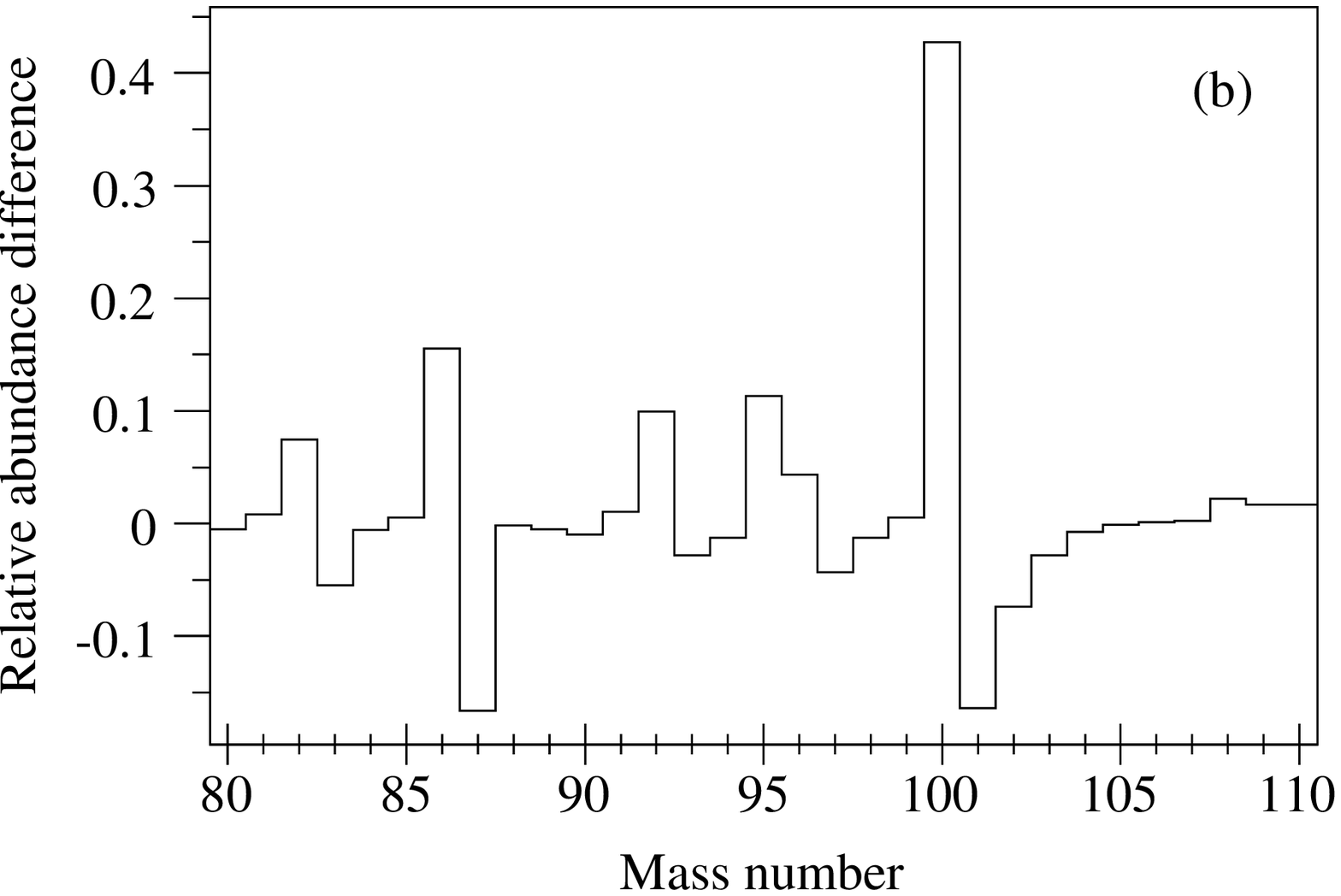}}

 \caption{(color online). (a) Abundance as a function of the mass number of the composition of X-ray burst ashes calculated with an X-ray burst model with (black) and without (red) $\beta p$ emission. (b) Relative abundance difference between the two calculations as a function of mass number.}
 \label{bpresults}
\end{figure}

The calculated composition of the burst ashes obtained without any $\beta p$ emission (this corresponds to previous calculations where this decay mode was neglected) was compared with calculations that took into account the experimentally-determined $\beta p$ branches, including the ones indicated in Fig.~\ref{chart} [see Fig.~\ref{bpresultsa}]. 
Noticeable effects occur for mass $A=83,87,93,97$, and in particular for mass $A=101$, because of the sizable $b_{\beta p}$ values of $^{83}$Zr, $^{87}$Mo, $^{93}$Pd, $^{97}$Cd, and $^{101}$Sn [see Fig.~14(b)]. However, these effects are not dramatic due to the small $b_{\beta p}$ values involved.
 In the simplest picture, $\beta p$ emission should only play a role late in the freezeout, when proton capture rates are slow owing to lower temperature or hydrogen exhaustion, as otherwise an emitted proton would tend to be recaptured. As shown in Fig.~14(b), the increase in $A=100$ yield was about four times larger than what one would expect from a simple transfer of $A=101$ nuclei to $A=100$ after freezeout due to the $\beta p$ branching. This indicates that $\beta p$ emission also plays a role during the rp process because of inefficient recapturing of emitted protons. This leads to an increase of $A=100$ abundance and a decrease in abundance of $A=101$--104 because of the reduced reaction flow towards heavier nuclei.


\section{Conclusions}
The $\beta$ decay of a range of neutron-deficient nuclei near $^{100}$Sn was studied. Most of these nuclei exhibit $\beta$-delayed proton emission. $b_{\beta p}$ values for $^{96}$Cd, $^{98}$In$^{g,m}$, $^{99}$In, $^{89}$Ru, $^{91,92}$Rh, $^{93}$Pd, and $^{95}$Ag were determined for the first time. $\beta p$ emission from $^{97}$Cd was used to disentangle ground and isomeric state decays \cite{Lor11}. The precision of $b_{\beta p}$ for $^{101}$Sn was improved by an order of magnitude, enabling the use of this quantity, together with the shape of the $\beta p$-energy spectrum, as a probe for shell-model calculations. However, to resolve the question of the unknown ground state spin of $^{101}$Sn, more precise mass measurements are necessary. The $\beta$-decay scheme of $^{93}$Pd was revised, and the energy of excited states in $^{95}$Rh were deduced with improved precision. In some cases, the uncertainties of half-lives were also improved.

The new half-lives and $b_{\beta p}$ data were used to update astrophysical rp-process network calculations. The $b_{\beta p}$ for $^{89}$Ru, $^{92}$Rh, $^{93}$Pd, $^{95}$Ag, $^{96,97}$Cd, $^{98,99}$In, and $^{101}$Sn provide a complete set of $b_{\beta p}$ data for the rp process between $A=92$--101, with the exception of $^{100}$Sn, which is not strongly populated in most rp-process models. The main outcome was the observation that the $b_{\beta p}$ values in this region tend to be small, resulting in limited influence on energy production and on the final abundance distribution. The one exception is the production of $A=100$ nuclei, which is enhanced by $\beta p$ emission from $^{101}$Sn, and which is now determined with reasonable precision. Earlier speculations that $\beta p$ emission may significantly alter the composition of heavy rp-process ashes is in fact not supported by these new results. The impact of the $\beta p$ emission on the production of $^{92}$Mo and $^{96}$Ru in the rp process was also addressed. Speculations of a large $b_{\beta p}$ of $^{93}$Pd seemed particularly intriguing in light of the pulsed rp process \cite{Boy98}, which underestimates the production of $^{92}$Mo and overestimates the production of $^{93}$Ru. The results reported here show that the relevant $b_{\beta p}$ are small and do not significantly affect rp-process nucleosynthesis in the $A=92$--98 region. Along with recent mass measurements \cite{Hae11, Tu11, Bre09}, the results presented here are an important step towards more reliable rp-process calculations in this mass region. With the ${\beta}$-decay properties largely pinned down, the remaining uncertainties arise from the unknown nuclear masses near the proton drip line and the untested proton capture rates in this entire region. 
\\
\\
We would like to thank the operations department of the NSCL for providing the high intensity $^{112}$Sn primary beam. This work is supported in part by NSF grants PHY02-16783, PHY-06-06007, PHY-1068217, and by Kakenhi grant 22$\cdot$00201.

\ctable[
caption={Half-lives and $b_{\beta p}$ measured in this work, compared to literature values when available.},
label = tab_results,
pos = b,
star
]{ll|c|ccc|ccc}{
\tnote[a]{Half-life inferred from $\beta$-decay data.}
\tnote[b]{Half-life inferred from weighted average of $\beta \gamma$ and $\beta p$-decay data.}
\tnote[c]{$\beta_{\beta p}$ measured assuming as half-life the weighted average 1.60(15)~s from Refs. \cite{Dea04,Kie01}.}
}{
\FL   
\hline
\vspace{-3mm}
&&&&&&\\
Nucleus& ~~J$^{+}$ & ~~detected~~ &\multicolumn{3}{c}{Half-life (s)}&\multicolumn{2}{c}{$b_{\beta p}$ (\%)}\NN
 \cmidrule(r){4-5}\cmidrule(r){6-7}
            &       &  protons    &This work & Literature &  Theory  & This work &~~ Literature  \NN
\hline
&&&&&&&\NN
$^{89}$Ru &    & 6 & 2.2(12) & 1.2(2)~\cite{Zha99} & 4.04~\cite{Bro97}&  3$^{+1.9}_{-1.7}$  &  \NN 
&&&&&&&\NN
$^{91}$Rh  &            &  12   &         & 1.47(22)~\cite{Dea04}& 1.68~\cite{Bro97}   & 1.3(5)\tmark[c] &    \NN
          &            &       &          & 1.7(2)~\cite{Kie01} &  &         &   \NN
&&&&&&&\NN
$^{92}$Rh$^{(g)}$ & ~6$^{+}$   &   1800  & 5.7(1)   & 4.6(25)~\cite{Dea04}&   &  1.9(1) &    \NN
$^{92}$Rh$^{(m)}$ & (2$^{+}$) &         &          & 0.5(2)~\cite{Dea04}  & 4.30~\cite{Bro97} &         &   \NN
&&&&&&&\NN
$^{93}$Pd &    & 1548 & 1.115(45) & 1.0(2)~\cite{Kie01}  &1.4 ~\cite{Bro97}   & 7.4(4) &$<$5~\cite{Schm00} \NN 
          &            &         &            & 0.9(2)~\cite{Schm00}&   &      &  \NN
          &            &         &            & 1.3(2)~\cite{Xu01}  &   &       &   \NN
&&&&&&&\NN
$^{95}$Pd &      &      & 7.5(5)\tmark[a] &           & 7.93~\cite{Bro97}& &  \NN 
&&&&&&&\NN
$^{95}$Ag &            &   1705  & 1.85(8)  & 2.0(1)~\cite{Schm94}  &1.85~\cite{Bro97}  &  2.5(3) &    \NN
          &            &         &          & 1.3(2)~\cite{Schm97} &  &         &   \NN
&&&&&&&\NN
$^{96}$Ag & (2$^{+}$)  &46858&4.395(85)& 4.40(6)~\cite{Bat03}&  11~\cite{Bro97}   & 6.5(8)& 8.5(15)~\cite{Bat03}\NN
          & (8$^{+}$)  &43540&6.8(10)  & 6.9(6)~\cite{Bat03} &    &  14(3) & 18(5)~\cite{Bat03}  \NN
&&&&&&&\NN
$^{96}$Cd &   &   12 & 1.03$^{+0.24}_{-0.21}$\tmark[a]~\cite{Baz08} &   & 2.18~\cite{Bro97}& 5.5(40)     &        \NN
&&&&&&&\NN
$^{97}$Cd &(9/2$^{+}$) &   4160  & 1.0(1)~\cite{Lor11}  & 2.8(6)~\cite{Schm97}& 1.16~\cite{Bro97} &11.8(20) &    \NN
          &          &         &                      &                     & 0.9, 1.1~\cite{Schm97}                   &         &  \NN
          &(25/2$^{+}$)&   7723   & 3.8(2)~\cite{Lor11}   &      &0.6~\cite{Schm97}  & 25(4)     &  \NN
&&&&&&&\NN
$^{98}$In$^{(g)}$&& 35000 &0.047(13)\tmark[a]&0.032$^{+32}_{-11}$~\cite{Kie01}    & 0.0181~\cite{Bro97} &5.5${^{+3}_{-2}}$ & \NN
  $^{98}$In$^{(m)}$&&       &  1.27(30)          & 1.2$^{+1.2}_{-0.4}$~\cite{Kie01}&& & 19.5(13)  &  \NN
&&&&&&&\NN
$^{99}$In &      &35000 &3.1(2)\tmark[a]&  3.0${^{+8}_{-7}}$~\cite{Kie01} & &   0.9(4)  & \NN
          &      &      &3.05(60)       &    &   &  & \NN
&&&&&&&\NN
$^{100}$In&  &   35000 & 5.7(3)\tmark[b] & 5.9(2)~\cite{Ple02} &&1.7(4)  &  1.6(3)~\cite{Ple02} \NN
          &  &         & 6.0(5)          & 6.1(9)~\cite{Sze95} &     &&  0.8(3)~\cite{Sze95}  \NN
          &  &         &                 & 6.7(7)~\cite{Sto02} &     &&   \NN
&&&&&&&\NN
$^{100}$Sn& &0  & 0.55$^{+0.70}_{-0.31}$\tmark[a]~\cite{Baz08} & 0.94$^{+0.54}_{-0.26}$~\cite{Baz08} & & $<$35 \NN
&&&&&&&\NN
$^{101}$Sn& &458  & 2.1(2) & 1.5(6)~\cite{Sto01}& &22(2)&14$^{+10}_{-6}$~\cite{Sto01}& \NN
          & &    &        & 1.9(3)~\cite{Kav07}&      &                          &    \NN
          & &     &        & 3(1)~\cite{Jan95}   &     &                          &    \LL
\hline
}

\newpage


\end{document}